\documentclass[preprint]{aastex}
\usepackage{graphicx}        % For eps figures, newer & more powerfull
\usepackage{natbib}         % For citations: redefine \cite commands
\usepackage{amssymb}        % useful mathemati cal symbols
\usepackage{amsmath}
\usepackage{color}           % For color text: \color command
\usepackage{url}             % For breaking URLs easily trough lines
\usepackage{epstopdf}
\usepackage{soul}
\providecommand\bnabla{\boldsymbol{\nabla}}
\providecommand\bcdot{\boldsymbol{\cdot}}
\newcommand\bu{{\boldsymbol{u}}}

\newcommand\be{{\boldsymbol{e}}}

\newcommand\bF{{\boldsymbol{F}}}
\newcommand\bx{{\boldsymbol{x}}}

\newcommand\bk{{\boldsymbol{k}}}

\renewcommand\Re{{\rm Re}}
\newcommand\Ri{{\rm Ri}}
\newcommand\Pe{{\rm Pe}}
\newcommand\p{\partial}

\begin{document}

\pagestyle{empty} %No headings for the first pages.

%% Title Page %%%%%%%%%%%%%%%%%%%%%%%%%%%%%%%%%%%%%%%%%%%%%%%
%% ==> Write your text here or include other files.

%% The simple version:
\title{Turbulent transport in a strongly stratified forced shear layer with thermal diffusion}

\author{Pascale Garaud, \\ Department of Applied Mathematics and Statistics, Baskin School of Engineering, \\University of California at Santa Cruz, 1156 High Street, Santa Cruz CA 95064. \\ Logithan Kulenthirarajah,\\
Institut de Recherche en Astrophysique et Planétologie (IRAP), 14, avenue Edouard Belin, 31400 Toulouse, France.
 }

%\date{} %%If commented, the current date is used.
\maketitle

% The nice version:
%\input{titlepage} %%You need a file 'titlepage.tex' for this.
%% ==> TeXnicCenter supplies a possible titlepage file
%% ==> with its templates (File | New from Template...).

%\begin{abstract}
\vspace{1cm}
\centerline{\bf Abstract} 
This work presents numerical results on the transport of heat and chemical species by shear-induced turbulence in strongly stratified but thermally diffusive environments. The shear instabilities driven in this regime are sometimes called  ``secular" shear instabilities, and can take place even when the gradient Richardson number of the flow (the square of the ratio of the buoyancy frequency to the shearing rate) is large, provided the P\'eclet number (the ratio of the thermal diffusion timescale to the turnover timescale of the turbulent eddies) is small. We have identified a set of simple criteria to determine whether these instabilities can take place or not. Generally speaking, we find that they may be relevant whenever the thermal diffusivity of the fluid is very large (typically larger than $10^{14}$cm$^2$/s), which is the case in the outer layers of high-mass stars ($M\ge 10 M_\odot$) for instance. Using a simple model setup in which the shear is forced by a spatially sinusoidal, constant-amplitude body-force, we have identified several regimes ranging from effectively unstratified to very strongly stratified, each with its own set of dynamical properties. Unless the system is in one of the two extreme regimes (effectively unstratified or completely stable), however, we find that (1) only about 10\% of the input power is used towards heat transport, while the remaining 90\% is viscously dissipated; (2) that the effective compositional mixing coefficient is well-approximated by the model of Zahn (1992), with $D \simeq 0.02 \kappa_T /J$ where $\kappa_T$ is the thermal diffusivity and $J$ is the gradient Richardson number. These results need to be confirmed, however, with simulations in different model setups and at higher effective Reynolds number.

\section{Introduction}
\label{sec:intro}

The continued progress in stellar spectroscopy, combined with the advent of asteroseismology, have opened new prospects for stellar astrophysics, providing more reliable ways of testing the accuracy of stellar evolution models. These challenge our understanding of the micro- and macro-physical processes that participate in all aspects of a star's life, from the deep interior to the surface. 
Today, the greatest sources of uncertainty in stellar modeling is arguably turbulent mixing in radiative regions. The need for non-canonical mixing (i.e. turbulent mixing of chemical species or angular momentum that is not related to convection or overshoot) in stellar evolution has long been recognized: despite the general agreement between 
theory and observations of stars, some discrepancies remain that can only be reconciled by invoking {\it additional} mixing in stellar radiation zones \citep[see the review by][for instance]{Pinsonneault97}. 
Examples of observational evidence for missing compositional mixing comes from surface abundances \citep[see for instance][]{Skumanich72,Fliegner96,Gratton00,SestitoRandich05,CharbonnelZahn07} and helioseismology \citep[e.g.][]{ElliottGough99} while that related to missing angular-momentum mixing comes for instance from asteroseismology of Red Giant Branch (RGB) stars \citep{Mosser12b,Eggenberger12,Deheuvelsetal14}.

Solving the ``missing mixing'' problem begins with identifying possible sources of hydrodynamical or magneto-hydrodynamical instabilities, determining under which condition they operate, and then quantifying the amount of heat, angular momentum, and compositional transport they induce. A commonly discussed mixing mechanism is shear-induced turbulence \citep[e.g.][]{Zahn1974,Schatzman77,EndalSofia78,Zahn92,Barnes99}. Shear -- in particular rotational shear -- is, to various levels, always present in a star. Whether shear-induced turbulence develops or not, and how much mixing it causes, depends on the amplitude of the viscous force and of the buoyancy force acting on the fluid, which both have a tendency to quench the instability.  In what follows, we now discuss in turn what is known about the conditions for which shear instabilities can develop in stars, which will lead us to discuss the effects of thermal diffusion, and then summarize some of the main shear-induced mixing models that have been proposed. 

\subsection{Under which conditions do shear instabilities occur? } 
\label{sec:crit}

The nature of the criterion for the onset of shear instabilities has long been the subject of debate, starting from the work of \citet{Richardson1920} who argued on energetic principles that shear-induced turbulence can only be sustained provided $J < J_c$, where $J = N^2/S^2$ is the local gradient Richardson number, $N$ is the local Brunt-V\"ais\"al\"a frequency (often called buoyancy frequency), $S = |d \bu/dr|$ is the local shearing rate of the flow field $\bu$ and $J_c$ is a universal constant which he argues must be of order one. Later, \citet{Miles61} and \citet{Howard61} formally proved that a necessary\footnote{but not always sufficient} condition for linear instability to occur in an inviscid and non-diffusive shear flow is that $J$ must drop below 1/4 somewhere in the fluid. Unfortunately, $J$ rarely ever drops below one in stellar interiors -- typical values are far greater, suggesting that the shear should always be stabilized by the stratification (both from linear theory, and on energetic principles). 

The effect of thermal diffusion, however, can relax this so-called Richardson criterion by damping the buoyancy force \citep{Townsend58,Dudis1974,Zahn1974,GageMiller74,Jones1977}. Indeed, in cases where the background temperature gradient within the star is the main source of stratification (by contrast with the possibility of chemical stratification which we ignore in this work), then the potential energy cost of overturning motions is reduced if the moving fluid has time to adjust thermally to its surroundings. Whether it does or not is typically measured by the eddy P\'eclet number 
\begin{equation}
\Pe_l = \frac{S l^2 }{\kappa_T}  \mbox{  ,}
\label{eq:Pel1}
\end{equation}
where $l$ is the vertical lengthscale of the eddy considered, and $\kappa_T$ is the thermal diffusivity. If $\Pe_l \ll 1$, then the motion is strongly diffusive, the eddy is always in thermal equilibrium with its surrounding, and the influence of thermal stratification effectively disappears -- at least on the scale $l$. 

While a strict yet simple and general linear stability criterion akin to the Miles-Howard theorem does not exist in the case of diffusive shear flows, a combination of mathematical and physical arguments backs the notion that thermal diffusion vastly increases the critical gradient Richardson number $J_c$ above which a flow is stable. Indeed, \citet{Townsend58} and \citet{Dudis1974} first performed a linear stability analysis of stratified atmospheric shear flows in the optically thin limit, where radiative losses play a similar role as thermal diffusion in damping the buoyancy force. In both cases, they found that thermal 
radiation begins to influence the stability of the fluid when $S t_{\rm cool}$ roughly drops below one (where $t_{\rm cool}$ is the thermal cooling timescale associated with the optically thin radiation law). The inviscid linear instability criterion is no longer $J   < J_c = 1/4$, but becomes $ J <  J_c \sim O(S t_{\rm cool})^{-1} $ instead, which implies that strongly stratified shear flows with $J \gg 1$ can still be destabilized provided $St_{\rm cool} \ll 1$. 

\citet{Zahn1974} used these results to conjecture on the stability of optically thick shear flows, which are much more relevant for stellar interiors. The equivalent cooling timescale $t_{\rm cool}$ for an eddy of size $l$ can be approximated by $t_{\rm cool} = l^2 / \kappa_T$, so the condition $S t_{\rm cool} < O(1) $ is equivalent to the low P\'eclet number condition discussed earlier, 
\begin{equation}
\Pe_l  < O(1)   \mbox{  ,}
\end{equation}
where $\Pe_l$ was defined in (\ref{eq:Pel1}). Proceeding with the analogy with the optically thin case, Zahn argued that the linear inviscid criterion for instability $ J S t_{\rm cool} < O(1)$ should therefore become
\begin{equation}
J \Pe_l  < (J\Pe)_c  \mbox{  ,}
\label{eq:jpel1}
\end{equation}
where $(J\Pe)_c$ is a constant of order unity. This new criterion clearly allows for instability at large $J$ provided $\Pe_l$ is small enough. 

By contrast with the optically thin limit, however, this criterion depends on the so-far-unspecified length scale $l$ of the turbulent eddies, and naively suggests that one could always pick $l$ small enough for (\ref{eq:jpel1}) to apply. In other words, shear instabilities should always be possible as long as there is an infinitesimal amount of thermal diffusion. In practice, however, this is of course not the case, since viscosity cannot be ignored on very small scales. \citet{Zahn1974} therefore argued that $l$ should be set equal to the smallest scale that is not stabilized by viscosity, $l_c$,  which satisfies $S l_c^2 /\nu = \Re_c$ where $\nu$ is the kinematic viscosity and the critical Reynolds number $\Re_c$ is a universal constant which Zahn suggests is of the order of a thousand. On that scale, $\Pe_l = \Pr  \Re_c$, where $\Pr  = \nu/\kappa_T$ is the Prandtl number and
the criterion for instability (\ref{eq:jpel1}) becomes
\begin{equation} 
 J \Pr \Re_c < (J\Pe)_c   \Leftrightarrow \frac{J \Pe}{(J\Pe)_c} <  \frac{\Re}{\Re_c}   \mbox{  ,}
 \label{eq:jzahn}
\end{equation}
where $\Pe$ and $\Re$ can now be defined according to any length scale $L$, instead of $l_c$ specifically. Note how while (\ref{eq:jpel1}) only addressed the question of the stability of a shear flow to perturbations of a given size $l$, (\ref{eq:jzahn}) more generally addresses the question of the stability of shear flows to perturbations at arbitrary lengthscales. This new criterion clearly allows for the development of instabilities for values of $J$ much larger than one provided the Prandtl number $\Pr$ is sufficiently low, and is sometimes referred to as a criterion for ``secular shear instabilities" \citep{EndalSofia78}. We note, however, that Zahn's criterion is not a formal mathematical result on either linear or nonlinear stability of stratified diffusive shear flows\footnote{If anything, Zahn's argument is necessarily one that pertains only to the nonlinear stability of the flow, since for small enough scales, the background shear will always be very nearly linear, and linear shear flows are inherently linearly stable.}. Rather, it should be viewed as a plausible heuristic criterion whose validity remains to be verified. Furthermore, given that this criterion is in principle only valid for low P\'eclet number flows, its domain of applicability also needs to be clarified (see below for more on this topic). 

Significant progress has been made in this direction in the past 20 years. \citet{Lignieres1999} and \citet{Lignieresetal1999} showed that the dynamics of low P\'eclet number flows in general can be studied with a reduced set of equations in which the temperature field is slaved to the velocity field \citep[see Section \ref{sec:model}, and see also][]{Thual}. In these equations, the Richardson number and the P\'eclet number always appear together as a product called the ``Richardson-P\'eclet" number hereafter. This formally shows that any criterion for stability or instability to shear at low P\'eclet number should involve the product of the 
Richardson number and the P\'eclet number rather than the Richardson number alone, which is indeed the case in (\ref{eq:jpel1}) and (\ref{eq:jzahn}) for instance. Using these so-called ``low P\'eclet number" (LPN) equations, \citet{Lignieresetal1999} then showed that the linear stability properties of a viscous stratified hyperbolic tangent shear layer do indeed satisfy a criterion of the kind written in (\ref{eq:jzahn}) \citep[see also][]{Jones1977}. First results in the nonlinear regime were obtained by \citet{PratLignieres13}, who showed numerically that a viscous stratified {\it linear} shear layer is nonlinearly unstable to finite amplitude perturbations that have a set lengthscale $l$ provided $J \Pe_l < (J\Pe)_c \simeq 0.426$, hence verifying (\ref{eq:jpel1}). Their experimental protocol (see below for more detail) however was unable to address the more general and more crucial question of the validity of (\ref{eq:jzahn}). 

More recently, \citet{Garaudal15} showed that any {\it periodic} stratified shear flow in the limit of asymptotically low P\'eclet number is formally stable to any initial perturbations (of arbitrary amplitude and shape) provided the Richardson-P\'eclet number is greater than a factor of order unity times the Reynolds number, which is equivalent to Zahn's criterion (\ref{eq:jzahn}) except with $\Re_c \sim O(1)$ instead of $O(10^3)$. This implies that the entire region $J \Pr < O(1)$ could in principle be subject to finite amplitude instabilities, potentially expanding the region of parameter space prone to shear instabilities even further compared with Zahn's predictions, albeit in the limit of small P\'eclet numbers. In practice, \citet{Garaudal15} found that only a subset of that region is nonlinearly unstable, but their results remain to be confirmed with more extensive simulations as well as with non-periodic flows. In summary, all of these studies point in the same direction, namely that Zahn's criterion given in (\ref{eq:jzahn}) appears to hold for low P\'eclet number flows both in the linear sense and in the nonlinear sense, although the actual values of $(J\Pe)_c$ and $\Re_c$ remain to be determined in each case, and depend on the global shape of the shear flow.

 \subsection{What is a low P\'eclet number flow?}
 
  The domain of validity of the LPN equations, and by proxy, the domain of applicability of criterion (\ref{eq:jzahn}), is clearly limited to low P\'eclet number flows. However, since one can define a number of different P\'eclet numbers depending on the length- and velocity-scales of interest,  it is worth asking {\it which} P\'eclet number has to be small for these equations to be valid. The original derivation of \citet{Lignieres1999} argues that one should use the r.m.s. velocity of the flow, $u_{\rm rms}$, together with the typical vertical scale of energy-bearing eddies $l_v$ to do so. In other words, the LPN equations and criterion (\ref{eq:jzahn}) should apply as long as the {\it turbulent} P\'eclet number 
 \begin{equation}
 \Pe_t \equiv \frac{u_{\rm rms} l_v }{ \kappa_T } \ll 1 \mbox{   .}
 \label{eq:Pel2}
 \end{equation} 
 
This seems to be confirmed by the numerical simulations of \citet{PratLignieres13} \citep[see also][]{PratLignieres14}, and of \citet{Garaudal15}. %, who both studied the effect of thermal diffusion on stratified shear flows using very different approaches \citep[see also][]{BruggenHillebrandt01}. 
As mentioned above, \citet{PratLignieres13} considered strictly linear stratified shear flows, which are known to be linearly stable \citep{Knobloch84}. Each one of their simulations has a well-defined background shearing rate $S$, and is initiated with finite amplitude perturbations that have a given velocity spectrum defining a dominant eddy size $l_v$, and a typically velocity scale $u_{\rm rms}$ (which together define their initial turbulent P\'eclet number $\Pe_t =  u_{\rm rms}l_v/\kappa_T$). By comparing the temporal evolution of the total kinetic energy in the various simulations, they were able to identify nonlinearly stable cases  (i.e. cases where the kinetic energy gradually decays to zero) from nonlinearly unstable ones (i.e. where the total kinetic energy increases), and thus measured numerically the critical threshold for nonlinear instability. They found that the critical Richardson-P\'eclet  number for the onset of instability as measured in simulations that use the LPN equations was an excellent match to those obtained using the full Naviers-Stokes equations in the limit where $\Pe_t$ was smaller than one, and that it is given by $(J\Pe)_c \simeq 0.426$ in that limit. In a related study, \citet{Garaudal15} studied the development of shear instabilities in sinusoidally forced stratified shear flows. They defined their P\'eclet number based on the macroscopic length scale $L$ and velocity $U_L$ of the laminar solution of the forced problem, $\Pe = U_L L/\kappa_T$. As for \citet{PratLignieres13}, they found that the critical Richardson-P\'eclet number for the onset of instability as measured in simulations that use the LPN equations was an excellent match to the one obtained with the full equations provided $\Pe < 1$. However, they also found that the domain of validity of the LPN equations extended to larger $\Pe$ as well, a result they attributed to the fact that the turbulent P\'eclet number $\Pe_t$  of the simulation could be smaller than one even if $\Pe$ was larger than one. 

The results of \citet{Garaudal15} illustrate that, while correct, Lignieres' criterion for the applicability of the LPN equations and of criterion (\ref{eq:jzahn}) ($\Pe_t < 1$) is not always practical for stellar evolution calculations  since it requires a priori knowledge of the typical eddy scale $l_v$ and velocity $u_{\rm rms}$ of the turbulent flow, but these can only be obtained a posteriori from full hydrodynamic simulations. Instead, it would be more desirable to have a criterion that predicts under which conditions the LPN equations will apply, based only on known global parameters. With this knowledge, we would then be able to determine fairly easily whether there are stars for which the low P\'eclet number limit is even relevant, and then apply Zahn's criterion more specifically to determine which stellar regions may be subject to secular shear instabilities at large Richardson number. 
 
 \subsection{How much mixing do shear instabilities cause?}
 
Knowing when secular shear instabilities are likely to occur is only half of the problem -- one also needs to quantify the transport of momentum, heat and composition induced by these instabilities, in order to make progress towards solving the missing mixing problem. \citet{Zahn1974} did not seem to address the question, and the first reference to the mixing rate of shear instabilities in the presence of strong diffusion can be found in the work of \citet{EndalSofia78} instead. They argued, as is commonly done, that the turbulent mixing coefficient $D$ can be estimated by multiplying a typical lengthscale $l$ to a typical velocity $u \sim Sl$, and proposed that $l$ be the {\it smallest} possible scale $l_c$ for which $Sl_c^2 / \nu = \Re_c$ (i.e. the same scale discussed earlier in the context of the secular shear instability criterion). This then yields
\begin{equation}
D \simeq Sl_c^2 \simeq  \nu \Re_c \mbox{  ,}
\end{equation}
which implies that $D$ should always be larger than $\nu$ by a factor of $\Re_c$ regardless of the level of stratification and regardless of the thermal diffusivity, as long as the criterion for secular shear instabilities is met. 
%l^2 = (nu Recrit / 
%v = l / T = Sl = S (nu Recrit / S)^1/2 = (nu Recrit S)^{1/2}
%vl = (nu Recrit S)^{1/2} (nu Recrit / S)^(1/2) = nu Recrit 

Later, \citet{Zahn92} argued that the relevant lengthscale $l$ chosen in the calculation of $D$ should instead be the {\it largest} possible scale for which $J \Pe_l$ is still in the nonlinearly unstable regime, in other words, one for which 
\begin{equation}
J \Pe_l = J \frac{S l^2 }{\kappa_T} = (J\Pe)_c \mbox{  ,}
\end{equation}
where $(J\Pe)_c$ is the same universal constant of order one discussed in the context of equation (\ref{eq:jpel1}) \citep[see][for a discussion of this formula]{PratLignieres13}. This then implies that the diffusion coefficient $D$ must take the form
\begin{equation}
D = \beta  Sl^2 = \beta  (J\Pe)_c \frac{\kappa_T}{J} = \nu \beta \frac{(J\Pe)_c}{J \Pr}  \mbox{  ,}
\end{equation}
where $\beta$ is another proportionality constant of order one. This shows that $D$ should simply be proportional to $\kappa_T / J$. If the flow is marginally stable to secular shear instabilities -- or in other words, if the flow just satisfies the stability criterion (\ref{eq:jzahn}) so that $J \Pr \simeq (J\Pe)_c/\Re_c$, then the estimates from \citet{EndalSofia78} and from \citet{Zahn92} are roughly equivalent. However, allowing for the possibility of flows that are significantly beyond marginal stability, that is, flows for which $J \Pr \ll \Re_c^{-1}$, then the mixing rate proposed by \citet{Zahn92}  can be orders of magnitudes larger than the one proposed by \citet{EndalSofia78}, and does depend both on the local shearing rate and stratification (through $J$) as well as on the local thermal diffusivity $\kappa_T$. Note that Zahn's prediction for $D$ must clearly fail in the strict limit where $J \rightarrow 0$. However, as discussed earlier, stellar shear layers are rarely, perhaps never, that weakly stratified, so the proposed coefficient is probably reasonable for all practical purposes. 

A first attempt to test numerically which of these two models for $D$ is more accurate was recently reported by \citet{PratLignieres13} and \citet{PratLignieres14}. Their experimental protocol, which was discussed above, led them to focus only on flows that were marginally stable to finite amplitude perturbations with a prescribed scale $l$. For these ``fixed lengthscale" marginal solutions, they found that $D \simeq 0.058 \kappa_T / J$ in the limit where $\Pe_l \le 1$, as predicted by \citet{Zahn92}. At a first glance, this appears to be a remarkable validation and calibration of Zahn's theory which also rules out the \citet{EndalSofia78} prescription. It is crucial to note, however, that \citet{PratLignieres13} effectively test Zahn's model precisely under the conditions for which it was derived (i.e. $J$ is selected to be at the marginal nonlinear stability threshold for the selected length scale $l$), so it is not entirely surprising to see that the model works well in that case. Whether these particular conditions would naturally arise in stars where the shear and the scale of the turbulent eddies are free to evolve remains to be determined, and therefore so is the validity of Zahn's model. 

In summarizing past work, we have therefore raised three questions. (1) What is the correct criterion to apply when trying to establish whether a particular stellar region is undergoing shear-induced mixing or not, taking into account the effect of thermal diffusion (but ignoring compositional stratification, at least for now) (2) Can one determine from first principles and without resorting to numerical calculations whether thermal diffusion will have a strong effect on the dynamics of stellar shear layers or not, and (3) What controls the turbulent mixing rates in stars where the shear layer and the resulting shear-induced turbulence are both free to evolve naturally? While the answer to question (1) has already been discussed fairly comprehensively in Section \ref{sec:crit} (and references therein), this paper focusses on questions (2) and (3). In Section \ref{sec:model} we present our model setup, and briefly discuss the LPN equations. Section \ref{sec:num} presents typical outcomes of the numerical experiments, both in the low P\'eclet number limit and in the high P\'eclet number limit, for weakly and strongly stratified shear flows. Section \ref{sec:globalresponse} addresses and answers question (2), namely under which simple conditions the LPN equations are valid approximations of the full Navier-Stokes equations. Section \ref{sec:stars} then discusses which stars are most likely to harbor regions where the LPN equations are valid, or equivalently, where secular shear instabilities might be relevant. Section \ref{sec:num2} looks more quantitatively at the available numerical data, and draws preliminary conclusions on the transport properties of secular shear instabilities thus partially addressing question (3). Section \ref{sec:ccl} finally discusses the implication of the results for astrophysical modeling, and suggests future avenues of investigation. 
 
\section{The model}
\label{sec:model}

We consider a region of a stellar radiation zone located around the radius $r_0$, with a local background density profile $\bar \rho(r) \simeq \rho_0 + (r-r_0) d\bar \rho/dr + \hdots $, and a local background temperature profile $\bar T(r) = T_0 + (r-r_0) d\bar T/ dr + \hdots $. The region also has a background adiabatic temperature gradient $d T_{\rm ad}/dr \simeq (T_0/P_0) \nabla_{\rm ad}$ where $P_0$ is the background pressure at $r = r_0$. 

We then consider a small domain around $r_0$, and model it using a Cartesian coordinate system with gravity defining the $z$ direction (so $z = r - r_0$, for instance). We assume that the domain is much thinner than a pressure scaleheight, and that any fluid flow is slow enough to use the Boussinesq approximation \citep{SpiegelVeronis1960}. As in the work of \citet{Garaudal15}, we assume that the shear is created by a body-force applied in the $x-$direction, whose amplitude varies in the $z-$direction but is otherwise constant with time: 
\begin{equation}
\bF = F(z) \be_x  \mbox{  .}
\end{equation}
This ``constant forcing" setup could model, for instance, the role of Euler's force in the generation of rotational shear by the very slow differential expansion and contraction of stellar layers. It could also model the effects of quasi-static tides in binary systems. For simplicity, we assume that $F(z) = F_0 \sin(kz)$ is spatially sinusoidal. While this may not be particularly realistic, the periodic nature of $F$ allows us to use a fully spectral code, which assumes that all the perturbations (to the temperature, velocity field, pressure) are triply-periodic. Section \ref{sec:ccl} discusses the potential pros and cons of using a sinusoidal force versus other possible ways of forcing the shear.

This system is described by the following dimensional equations (which neglect the role of compositional stratification): 
\begin{eqnarray}
&& \frac{\p \bu}{\p t} +  \bu \bcdot \bnabla \bu = - \frac{1}{\rho_0} \nabla  p + \alpha g T \be_z + \nu \bnabla^2 \bu +\frac{F_0}{\rho_0} \sin(kz) \be_x  \mbox{  ,} \nonumber  \\
&& \bnabla \bcdot  \bu = 0 \mbox{  ,} \nonumber \\
&& \frac{\p T}{\p t} + \bu \cdot \nabla T + \left(\frac{d\bar T}{dr}  - \frac{d T_{\rm ad}}{dr}\right) w = \kappa_T \nabla^2 T \mbox{  ,}
\label{eq:originalforcing}
\end{eqnarray}
where $p$ and $T$ are the pressure and temperature perturbations away from the means $P_0$ and $T_0$, and $\bu = (u,v,w)$ is the velocity field. The diffusivities $\kappa_T$ and $\nu$, as well as the thermal expansion coefficient $\alpha = - \rho_0^{-1} (\partial \rho/\partial T)$ and the local gravity $g$ are all assumed to be constant. 

As discussed by \citet{Garaudal15}, these equations have a laminar solution with no temperature fluctuations, where the velocity field is given by: 
\begin{equation}
\bu_L = \frac{F_0}{k^2 \nu \rho_0} \sin(kz) \be_x \equiv U_L \sin(kz) \be_x   \mbox{  .}
\label{lamsoldimensional}
\end{equation}
The amplitude and lengthscale of this velocity field can be used to define a new unit system where 
\begin{eqnarray}
&&[u] =U_L  = \frac{F_0}{k^2 \nu \rho_0}   \mbox{ is the unit velocity,} \nonumber  \\
&&[l] = k^{-1}  \mbox{ is the unit length,}  \nonumber \\ 
&&[t] = \frac{[l] }{ [u] } = \frac{k \nu \rho_0}{F_0}   \mbox{ is the unit time,} \nonumber \\  
&&[T] = k^{-1} \left|\frac{d\bar T}{dr}  - \frac{d T_{\rm ad}}{dr}\right|   \mbox{ is the unit temperature,}
\label{eq:unitslaminar}
\end{eqnarray}
in which case the non-dimensionalized equations become :
\begin{eqnarray}
&& \frac{\p \hat \bu}{\p t} +  \hat \bu \bcdot \bnabla \hat \bu = -  \bnabla  \hat p + \Ri \, \hat T \be_z + \frac{1}{\Re} \nabla^2 (\hat \bu- \sin(z) \be_x)  \mbox{  ,} \label{eq:originallaminar1}\\
&& \bnabla \bcdot \hat \bu = 0 \mbox{  ,}\\
&& \frac{\p \hat T}{\p t} + \hat \bu \cdot \bnabla \hat T +\hat w = \frac{1}{\Pe} \nabla^2 \hat T\mbox{  ,}
\label{eq:originallaminar}
\end{eqnarray}
where $\hat \bu$, $\hat p$ and $\hat T$ are the non-dimensional velocity, pressure and temperature fields, where the differential operators as well as the independent variables have been implicitly non-dimensionalized as well, and where
\begin{eqnarray}
\Re = \frac{U_L}{k\nu} = \frac{F_0}{\rho_0 \nu^2 k^3} \, , \nonumber \\
\Ri = \frac{N^2}{U_L^2 k^2} = \frac{\alpha g \left|\frac{d\bar T}{dr}  - \frac{d T_{\rm ad}}{dr}\right| \rho_0^2 \nu^2 k^2}{F_0^2} \, , \nonumber \\
 \Pe = \frac{U_L}{k\kappa_T} =  \frac{F_0}{\rho_0 \nu k^3 \kappa_T}   \mbox{  .}
\end{eqnarray} 
The laminar solution (\ref{lamsoldimensional}) now takes the dimensionless form $\hat \bu_L=\sin(z) \, \be_x$. It is worth bearing in mind that, being based on the hypothetical laminar solution of the equations, the Reynolds, P\'eclet and Richardson numbers just defined are not necessarily relevant to the dynamics of the turbulent solution. 

%\begin{eqnarray}
%&& [l] = k^{-1} \mbox{ as the unit distance,    }  [F] = F_0   \mbox{ as the unit force per unit volume,}  \\ 
%&& [T] = |T_{0z}-T_{0z}^{\rm ad}| k^{-1} \mbox{ as the unit temperature,} \nonumber \\
%&& [t] =  \left(\frac{\rho_0}{k F_0}\right)^{1/2} \mbox{ as the unit time, and }  [u] = \left(\frac{F_0}{k \rho_0}\right)^{1/2} \mbox{ as the unit velocity}  \nonumber
%\end{eqnarray}
%where $F_0$ is the dimensional amplitude of the force, $k$ is its spatial wavenumber, and  $T_{0z}-T_{0z}^{\rm ad}$ is the difference between the assumed background temperature gradient, and the local adiabatic temperature gradient. 
%%where the constants $\rho_0$, $\alpha$, $g$, $\nu$ and $\kappa_T$ are the mean density of the region considered, the coefficient of thermal expansion, the local gravity, the viscosity and the thermal diffusivity respectively.  

As discussed in Section \ref{sec:intro}, we will specifically be interested in low P\'eclet number flows, for which thermal diffusion is significant. According to \citet{Lignieres1999}, in this limit the 
temperature fluctuations are slaved to the vertical velocity fluctuations as 
\begin{equation}
\hat w = \frac{1}{\Pe} \nabla^2 \hat T \mbox{  ,}
\label{eq:slave}
\end{equation}
and the flow dynamics can be modeled using the reduced low P\'eclet number (LPN) momentum equation: 
\begin{eqnarray}
\frac{\p \hat \bu}{\p t} + \hat \bu \bcdot \bnabla \hat \bu = - \nabla  \hat p + \Ri \Pe \nabla^{-2} \hat w \be_z +  \frac{1}{\Re} \nabla^2 (\hat \bu- \sin(z) \be_x)   \mbox{  ,}
&& \bnabla \bcdot  \hat \bu = 0  \mbox{  .} 
\label{eq:lowPeT}
\end{eqnarray}
There are two clear advantages in using this system compared with the standard equations at low P\'eclet number \citep{Lignieres1999}: on the one hand, there is one less variable, and therefore one less equation to solve for, and on the other hand this reduced asymptotic system bypasses the complications that may arise from having to follow the evolution of several fields that evolve on vastly different timescales (i.e. the stiffness problem). Meanwhile, the inverse Laplacian is dealt with trivially in our triply periodic spectral code. Finally, note that, as discussed by \citet{Garaudal15}, the LPN equations implicitly assume that the horizontally-averaged temperature profile must always be the same as the background profile, which is not the case for the full equations. We shall compare in this work the predictions of the standard equations and of the LPN equations in more detail, using Direct Numerical Simulations. 

\section{Numerical simulations}
\label{sec:num}

\subsection{The numerical model}

In the majority of the simulations discussed in this paper, the set of equations (\ref{eq:originalforcing}) is solved in a triply-periodic domain of size $L_x = 10\pi$, $L_y = 4\pi$ and $L_z = 2\pi$, using the pseudo-spectral code originally developed by S. Stellmach \citep{Traxleretal2011b,Stellmachetal2011} and modified for the purposes of this work to include the body forcing. Simulations are either initiated with small random fluctuations in the velocity field, or started from the results of a previous run at different parameter values. Table \ref{table1} shows a record of all simulations run in this format, together with some of the mean properties of the turbulent solutions. We have also modified the code to solve instead the LPN momentum equation (\ref{eq:lowPeT}) together with the continuity equation, and have run a number of simulations with this new setup in a slightly smaller domain of size $L_x = 10\pi$, $L_y = 2\pi$ and $L_z = 2\pi$. The latter are summarized in Table \ref{table2a}. Note that the onset of instability in these simulations was already discussed by \citet{Garaudal15}, although we have also added new ones at larger values of the Richardson-P\'eclet number.

\begin{table}[]
	\caption{Summary of the main results for all the runs with the standard equations. All units and input parameters are defined based on the forcing, as discussed in Section \ref{sec:forcingpars}. All runs have $\Re_F=100$ (equivalently, $\Re = 10^4$), and a resolution of $960 \times 384 \times 192$ effective mesh points (equivalently, $320 \times 128 \times 64$ Fourier modes). The first column reports $\Pe_F = \Re^{-1/2} \Pe$, the second column reports $\Ri_F = \Re\Ri $. The third column is the time average of the r.m.s. velocity of the flow defined in (\ref{eq:urmsinst}), the fourth column is the time average of the vertical scaleheight defined in (\ref{eq:lv1}), the fifth column is (minus) the time-averaged and volume-averaged turbulent heat flux, the sixth column is (minus) the time-averaged and volume-averaged turbulent compositional flux, and the seventh column is the time- and volume-averaged input power (see Section \ref{sec:energy}).  }
	\label{table1}
\centering
\vspace{0.3cm}
\begin{tabular}{ccccccc}
	\tableline
		$\Pe_F$  & $\Ri_F$ & $\breve{u}_{\rm rms}    $    & $\breve{l}_v$         & $-\langle \breve w \breve T\rangle  $                &$ -\langle \breve w \breve C \rangle        $              &   $\langle \breve \bu \cdot \breve \bF \rangle$          \\
\tableline
		0.1 & $10^{-3}$ & 2.21 $\pm$ 0.06  & 0.791 $\pm$ 0.01 & 1.48 $\pm$ 0.2     & 12.7 $\pm$ 2   & 0.369 $\pm$ 0.04 \\
		0.1 & $10^{-2}$ & 2.2 $\pm$ 0.05   &  0.802 $\pm$ 0.01&  1.21 $\pm$ 0.2     & 10.1 $\pm$ 1   & 0.370 $\pm$ 0.03 \\
		0.1 & $10^{-1}$ & 2.4 $\pm$ 0.05  & 0.78 $\pm$ 0.017  & 0.48 $\pm$ 0.1     & 3.34 $\pm$ 0.6    & 0.624 $\pm$ 0.04 \\
		0.1 & 1 & 2.46 $\pm$ 0.09  & 0.7 $\pm$ 0.016  & 0.1 $\pm$ 0.03     & 0.94 $\pm$ 0.1    & 0.908 $\pm$ 0.07 \\
		0.1 & 10 & 3.39 $\pm$ 0.04  & 0.6 $\pm$ 0.006  & 0.01 $\pm$ 0.001   & 0.28 $\pm$ 0.01   & 1.86 $\pm$ 0.04 \\
		0.1 & $10^{2}$ & 5.12 $\pm$ 0.02  & 0.494 $\pm$ 0.003  & 0.004 $\pm$ 0.002   & 0.12 $\pm$ 0.003   & 3.31 $\pm$ 0.02 \\
		0.1 & $10^{3}$ & 11.3 $\pm$ 0.04  & 0.355 $\pm$ 0.002 & $9.6 \times 10^{-4}$ $\pm$ $4 \times 10^{-5}$ & 0.03 $\pm$ 0.001   & 7.86 $\pm$ 0.03 \\
		&        &                   &                     &                   &                   &                  \\
		1   & $10^{-4}$ & 2.21 $\pm$ 0.06  & 0.796 $\pm$ 0.01 & 8.5 $\pm$ 2      & 11.8 $\pm$ 3    & 0.366 $\pm$ 0.06 \\
		1   & $10^{-3}$& 2.20 $\pm$ 0.06  & 0.797 $\pm$ 0.01 & 7.81 $\pm$ 1     & 10.7 $\pm$ 2    & 0.368 $\pm$ 0.04 \\
		1   & $10^{-2}$ & 2.40 $\pm$ 0.06  & 0.775 $\pm$ 0.02  & 2.87 $\pm$ 0.5     & 3.67 $\pm$ 0.6   & 0.613 $\pm$ 0.04 \\
		1   & $10^{-1}$ & 2.30 $\pm$ 0.05   & 0.775 $\pm$ 0.02 & 1.59 $\pm$ 0.2     & 1.71 $\pm$ 0.2   & 0.7 $\pm$ 0.05  \\
		1   & $1$ & 2.92 $\pm$ 0.1 & 0.674 $\pm$ 0.02   & 0.25 $\pm$ 0.06     & 0.39 $\pm$ 0.05    & 1.42 $\pm$ 0.1  \\
		1   & 10 & 4.90 $\pm$ 0.03  & 0.518 $\pm$ 0.005 & 0.04 $\pm$ 0.002    & 0.125 $\pm$ 0.004  & 3.13 $\pm$ 0.03  \\
		1   & $10^{2}$ & 11.0 $\pm$ 0.04 & 0.384 $\pm$ 0.003 & 0.01 $\pm$ $3 \times 10^{-4}$   & 0.03 $\pm$ 0.001   & 7.61 $\pm$ 0.03  \\
		&        &                   &                     &                   &                   &                  \\
		10  & $10^{-4}$ & 2.23 $\pm$ 0.06 & 0.8 $\pm$ 0.01   & 14.3 $\pm$ 2    & 13.8 $\pm$ 2   & 0.367 $\pm$ 0.03 \\
		10  & $10^{-3}$ & 2.23 $\pm$ 0.09  & 0.8 $\pm$ 0.01   & 10.3 $\pm$ 1    & 10.0 $\pm$ 1   & 0.376 $\pm$ 0.04 \\
		10  & $10^{-2}$ & 2.26 $\pm$ 0.1  & 0.819 $\pm$ 0.02 & 7.83 $\pm$ 1    & 7.23 $\pm$ 1     & 0.446 $\pm$ 0.04 \\
		10  & $10^{-1}$ & 2.35 $\pm$ 0.09  & 0.762 $\pm$ 0.02  & 1.35 $\pm$ 0.3     & 1.32 $\pm$ 0.2    & 0.741 $\pm$ 0.07 \\
		10  & 1 & 2.74 $\pm$ 0.2  & 0.738 $\pm$ 0.03  & 0.22 $\pm$ 0.8     & 0.21 $\pm$ 0.9    & 0.86 $\pm$ 0.3  \\
		10  & 10 & 6.05 $\pm$ 1  & 0.750 $\pm$ 0.1   & 0.18 $\pm$ 0.4    & 0.14 $\pm$ 0.3   & 3.14 $\pm$ 1  \\
		10  & $10^{2}$ & 13.9 $\pm$ 3 & 0.893 $\pm$ 0.1  & 0.07 $\pm$ 0.3     & 0.05 $\pm$ 0.3    & 9.36 $\pm$ 3 
	\end{tabular}
\end{table}

\begin{table}[]
	\caption{Summary of the main results for all the runs with the LPN equations. All units and input parameters are defined based on the forcing, as discussed in Section \ref{sec:forcingpars}. The first column reports $\Re_F = \Re^{1/2}$, the second column reports $\Ri_F \Pe_F= \Re^{1/2} \Ri \Pe $. The third column is the number of equivalent mesh points in each direction. The fourth column is the time average of the r.m.s. velocity of the flow defined in (\ref{eq:urmsinst}). The fifth column is the time average of the vertical scaleheight defined in (\ref{eq:lv1}). The sixth column is (minus) the time-averaged and volume-averaged turbulent heat flux divided by $\Pe_F$ (see Sections \ref{sec:energy}). The seventh column is the mean shear at $z = \pi$ (see Section \ref{sec:mean}).}
	\label{table2a}
	\centering
	\vspace{0.3cm}
	\resizebox{\linewidth}{!}{\begin{tabular}{ccccccc}
		\tableline
		$\Re_F$ & $\Ri_F \Pe_F$  & $ N_x, N_y, N_z$  & $ \breve u_{\rm rms}$        & $ \breve l_v$                        & $ - \langle \breve w \breve T \rangle / \Pe_F$                 & $ d\breve{\bar{u}}/dz  (z=  \pi) $                         \\
			\tableline
		10     & $ 10^{-3}$     & 480, 96, 96    & 2.40 $ \pm$ 0.008    & 0.991 $ \pm$ 0.002             & 23.7 $ \pm$ 0.351                                 & 0.756 $ \pm$ 0.1              \\
		10     & $ 10^{-2}$     & 480, 96, 96    & 2.06 $ \pm$ 0.002  & 0.996 $ \pm$ 3$\times 10^{-5}$  & 16.5 $ \pm$ 1.20 $\times 10^{-2}$                   & 0.92 $ \pm$ 1$\times 10^{-5}$  \\
		10     & $ 10^{-1}$     & 480, 96, 96    & 2.12 $ \pm$ 0.2    & 0.944 $ \pm$ 0.03              & 1.84 $ \pm$ 0.9                                   & 1.81 $ \pm$ 1                 \\
		10     & 1          & 480, 96, 96    & 3.74 $ \pm$ 0.4    & 0.982 $ \pm$ 0.009             & 0.943 $ \pm$ 0.504                                & 4 $ \pm$ 2                    \\
		10     & 1.5        & 480, 96, 96    & 5.16 $ \pm$ 0.0001 & 0.991 $ \pm$ 3$\times 10^{-6}$  & 0.568 $ \pm$ 9.00$\times 10^{-5}$                  & 6.67 $ \pm$ 1$\times 10^{-5}$  \\
		10     & 2          & 480, 96, 96    & 6.40 $ \pm$ 0.005   & 0.993 $ \pm$ 3$\times 10^{-5}$  & 0.193 $ \pm$ 4.00$\times 10^{-3}$                  & 8.7 $ \pm$ 0.1                \\
		10     & 2.2        & 480, 96, 96    & 6.85 $ \pm$ 0.001  & 0.994 $ \pm$ 1$\times 10^{-6}$  & $ 6.30\times 10^{-3} \pm$ 0.0003  & 9.6 $ \pm$ 0.1                \\
		&            &                &                 &                             &                                                &                            \\
		33.3     & $ 10^{-3}$     & 480, 96, 96    & 2.18 $ \pm$ 0.1    & 0.901 $ \pm$ 0.02              & 13.1 $ \pm$ 4                                   & 0.86 $ \pm$ 0.3               \\
		33.3    & $ 10^{-2}$     & 480, 96, 96    & 2.18 $ \pm$ 0.1    & 0.910 $ \pm$ 0.02               & 9.53 $ \pm$ 3                                    & 1.05 $ \pm$ 0.5               \\
		33.3     & $ 10^{-1}$     & 480, 96, 96    & 2.34 $ \pm$ 0.1    & 0.850 $ \pm$ 0.02               & 1.64 $ \pm$ 0.6                                  & 1.83 $ \pm$ 0.9               \\
		33.3     & 1          & 720, 144, 144  & 3.42 $ \pm$ 0.1    & 0.731 $ \pm$ 0.01              & 0.202 $ \pm$ 0.04                                 & 3.5 $ \pm$ 0.8                \\
		33.3     & 3          & 720, 144, 144  & 4.25 $ \pm$ 0.1    & 0.665 $ \pm$ 0.01              & $9.10\times 10^{-2}$  $\pm$ 0.01                                 & 4.67 $ \pm$ 0.5               \\
		33.3     & 6          & 720, 144, 144  & 5.42 $ \pm$ 0.3    & 0.617 $ \pm$ 0.01              & $5.80\times 10^{-2}$ $\pm$ 0.02                                 & 6.2 $ \pm$ 1                  \\
		33.3    & 7          & 720, 144, 144  & 16.1 $ \pm$ 7      & 0.745 $ \pm$ 0.2               & $6.20\times 10^{-2}$ $\pm$ 0.3                                  & 18 $ \pm$ 10                  \\
		33.3     & 8          & 720, 144, 144  & 23.5 $ \pm$ 0.01   & 0.992 $ \pm$ 1$\times 10^{-7}$  & $7.31\times 10^{-3}$ $\pm 1\times 10^{-5}$                   & 33 $ \pm$ 0.001               \\

\end{tabular}
}
\end{table}

\begin{table}[]
	\centering
	\caption{Continued from Table 2.}
	\label{table2b}
	\centering
	\vspace{0.3cm}
	\resizebox{\linewidth}{!}{\begin{tabular}{ccccccc}
			\tableline
		$\Re_F$ & $\Ri_F \Pe_F$  & $ N_x, N_y, N_z$  & $ \breve u_{\rm rms}$        & $ \breve l_v$                        & $ - \langle \breve w \breve T \rangle / \Pe_F$                 & $ d\breve{\bar{u}}/dz  (z = \pi) $                         \\
		\tableline
		50     & $ 10^{-3}$     & 480, 96, 96    & 2.19 $ \pm$ 0.1    & 0.857 $ \pm$ 0.02              & 11.7 $ \pm$ 3                                     & 0.841 $ \pm$ 0.3              \\
		50     & $ 10^{-2}$     & 480, 96, 96    & 2.36 $ \pm$ 0.1    & 0.845 $ \pm$ 0.02              & 5.68 $ \pm$ 2                                     & 1.28 $ \pm$ 0.4               \\
		50     & $ 10^{-1}$     & 480, 96, 96    & 2.42 $ \pm$ 0.1    & 0.798 $ \pm$ 0.02              & 1.57 $ \pm$ 0.5                                   & 1.86 $ \pm$ 0.5               \\
		50     & 1          & 480, 96, 96    & 3.30 $ \pm$ 0.1     & 0.678 $ \pm$ 0.01              & 0.17 $ \pm$ 0.03                                  & 3.33 $ \pm$ 1                 \\
		50     & 3          & 720, 144, 144  & 3.99 $ \pm$ 0.1    & 0.627 $ \pm$ 0.009             & $ 8.89\times 10^{-2} \pm $ 0.008     & 4.25 $ \pm$ 0.7               \\
		50     & 6          & 720, 144, 144  & 4.73 $ \pm$ 0.1    & 0.586 $ \pm$ 0.008             & $ 6.00\times 10^{-2} \pm$ 0.005     & 5.32 $ \pm$ 1                 \\
		50     & 10         & 720, 144, 144  & 5.58 $ \pm$ 0.1    & 0.552 $ \pm$ 0.007             & $ 4.40\times 10^{-2} \pm$ 0.004     & 6.4 $ \pm$ 1                  \\
		50     & 15         & 720, 144, 144  & 6.51 $ \pm$ 0.1    & 0.522 $ \pm$ 0.006             & $ 3.30\times 10^{-2} \pm$ 0.004     & 7.6 $ \pm$ 1                  \\
		50     & 20         & 720, 144, 144  & 7.34 $ \pm$ 0.2    & 0.500 $ \pm$ 0.006               & $ 2.70\times 10^{-2} \pm$ 0.005    & 8.5 $ \pm$ 1                  \\
		50     & 25         & 720, 144, 144  & 10.9 $ \pm$ 0.1    & 0.487 $ \pm$ 0.007             & $ 2.70\times 10^{-2} \pm$ 0.003    & 9.47 $ \pm$ 1                 \\
		50     & 30         & 720, 144, 144  & 11.8 $ \pm$ 0.1    & 0.473 $ \pm$ 0.006             & $ 2.26\times 10^{-2} \pm$ 0.02     & 10.2 $ \pm$ 1                 \\
		50     & 35         & 720, 144, 144  & 12.5 $ \pm$ 0.1    & 0.46 $ \pm$ 0.006              & $ 1.97\times 10^{-2} \pm$ 0.002     & 11 $ \pm$ 1                   \\
		50     & 40         & 720, 144, 144  & 13.3 $ \pm$ 0.1    & 0.452 $ \pm$ 0.006             & $ 1.73\times 10^{-2} \pm$ 0.002    & 11.5 $ \pm$ 1                 \\
		50     & 45         & 720, 144, 144  & 14 $ \pm$ 0.2      & 0.442 $ \pm$ 0.006             & $ 1.51\times 10^{-2} \pm$ 0.004     & 12.4 $ \pm$ 1                 \\
		&            &                &                 &                             &                                                &                            \\
		100    & $ 10^{-4}$     & 960, 162, 162  & 2.27 $ \pm$ 0.07   & 0.794 $ \pm$ 0.02              & 14.8 $ \pm$ 4                                     & 0.83 $ \pm$ 2                 \\
		100    & $ 10^{-3}$     & 960, 162, 162  & 2.27 $ \pm$ 0.1    & 0.796 $ \pm$ 0.02              & 11.9 $ \pm$ 3                                     & 0.86 $ \pm$ 0.4               \\
		100    & $ 10^{-2}$     & 960, 162, 162  & 2.39 $ \pm$ 0.1    & 0.839 $ \pm$ 0.02              & 13 $ \pm$ 2                                       & 1.1 $ \pm$ 0.4                \\
		100    & $ 10^{-1}$     & 960, 162, 162  & 2.42 $ \pm$ 0.1    & 0.740 $ \pm$ 0.02               & 1.73 $ \pm$ 0.5                                   & 1.84 $ \pm$ 0.6               \\
		100    & 1          & 960, 162, 162  & 3.41 $ \pm$ 0.05   & 0.601 $ \pm$ 0.01              & 0.148 $ \pm$ 0.02                                 & 3.5 $ \pm$ 0.6                \\
		100    & 10         & 960, 162, 162  & 5.11 $ \pm$ 0.04   & 0.494 $ \pm$ 0.01              & $ 4.00\times 10^{-2} \pm$ 0.03                & 5.78 $ \pm$ 0.7               \\
		100    & 50         & 960, 162, 162  & 8.51 $ \pm$ 0.03   & 0.395 $ \pm$ 0.003             & $ 1.50\times 10^{-2} \pm$ 0.0005    & 10 $ \pm$ 1                   \\
		100    & 100        & 960, 162, 162  & 11.3 $ \pm$ 0.05   & 0.354 $ \pm$ 0.03              & $ 9.50\times 10^{-3} \pm $ 0.0004     & 13.1 $ \pm$ 1                 \\
		100    & 120        & 960, 162, 162  & 12.3 $ \pm$ 0.07   & 0.343 $ \pm$ 0.003             & $ 8.30\times 10^{-3} \pm $ 0.0005     & 14.1 $ \pm$ 1                 \\
		100    & 130        & 960, 162, 162  & 12.8 $ \pm$ 0.06   & 0.338 $ \pm$ 0.002             & $ 7.80\times 10^{-3} \pm $ 0.0004     & 14.6 $ \pm$ 1                 \\
		100    & 140        & 960, 162, 162  & 13.3 $ \pm$ 0.08   & 0.334 $ \pm$ 0.003             & $ 7.40\times 10^{-3} \pm $ 0.0005     & 15 $ \pm$ 1                   \\
		100    & 150        & 960, 162, 162  & 13.7 $ \pm$ 0.1    & 0.330 $ \pm$ 0.002              & $ 7.00\times 10^{-3} \pm $ 0.0006     & 15.7 $ \pm$ 0.9               \\
		100    & 160        & 960, 162, 162  & 14.5 $ \pm$ 0.7    & 0.326 $ \pm$ 0.003             & $ 6.70\times 10^{-3} \pm $ 0.003     & 16.1 $ \pm$ 1                 \\
		100    & 170        & 960, 162, 162  & 20.8 $ \pm$ 0.04   & 0.325 $ \pm$ 0.002             & $ 7.40\times 10^{-3} \pm $ 0.0003     & 17.5 $ \pm$ 1                 \\
		100    & 180        & 960, 162, 162  & 21.3 $ \pm$ 0.03   & 0.320 $ \pm$ 0.002              & $ 7.00\times 10^{-3}\pm$ 0.0003    & 17.6$ \pm$ 1 
	               
		\end{tabular}
	}
	
		\end{table}

\subsection{Typical results at low P\'eclet number}
\label{sec:qualreslowPe}

We first look at typical results in the limit of low P\'eclet numbers, for $\Re = 10^4$ (our largest $\Re$ value) except where specifically mentioned. As we shall demonstrate in Section \ref{sec:globalresponse} the behavior of all runs with the standard equations at $\Re = 10^4$ and $\Pe \le10$ is actually statistically indistinguishable from that of the LPN equations at the same $\Re$.  In all unstable cases, after a short transient period, the flow settles into a well-defined statistically stationary state. The behavior of the solutions in that state can loosely be classified into three categories: those with $\Ri \rightarrow 0$, which are all very similar to the unstratified limit $\Ri = 0$, those where $\Ri$ is somewhat larger, for which the effects of stratification becomes important, and those where $\Ri$ exceeds unity, which are well into the linearly stable region of parameter space, and for which new dynamics emerge. %In terms of the parameter $\Ri\Pe$, the first transition occurs roughly at $\Ri\Pe \sim 10^{-4}$ regardless of $\Re$ (see more on this transition in Section \ref{sec:globalresponse}). The second transition happens at a critical value of $\Ri\Pe$ that appears to depend on the Reynolds number (see more on these runs in Section \ref{sec:num2}). 

\subsubsection{The nearly unstratified limit} 

\begin{figure}[h]
  \centerline{\includegraphics[width=0.7\textwidth]{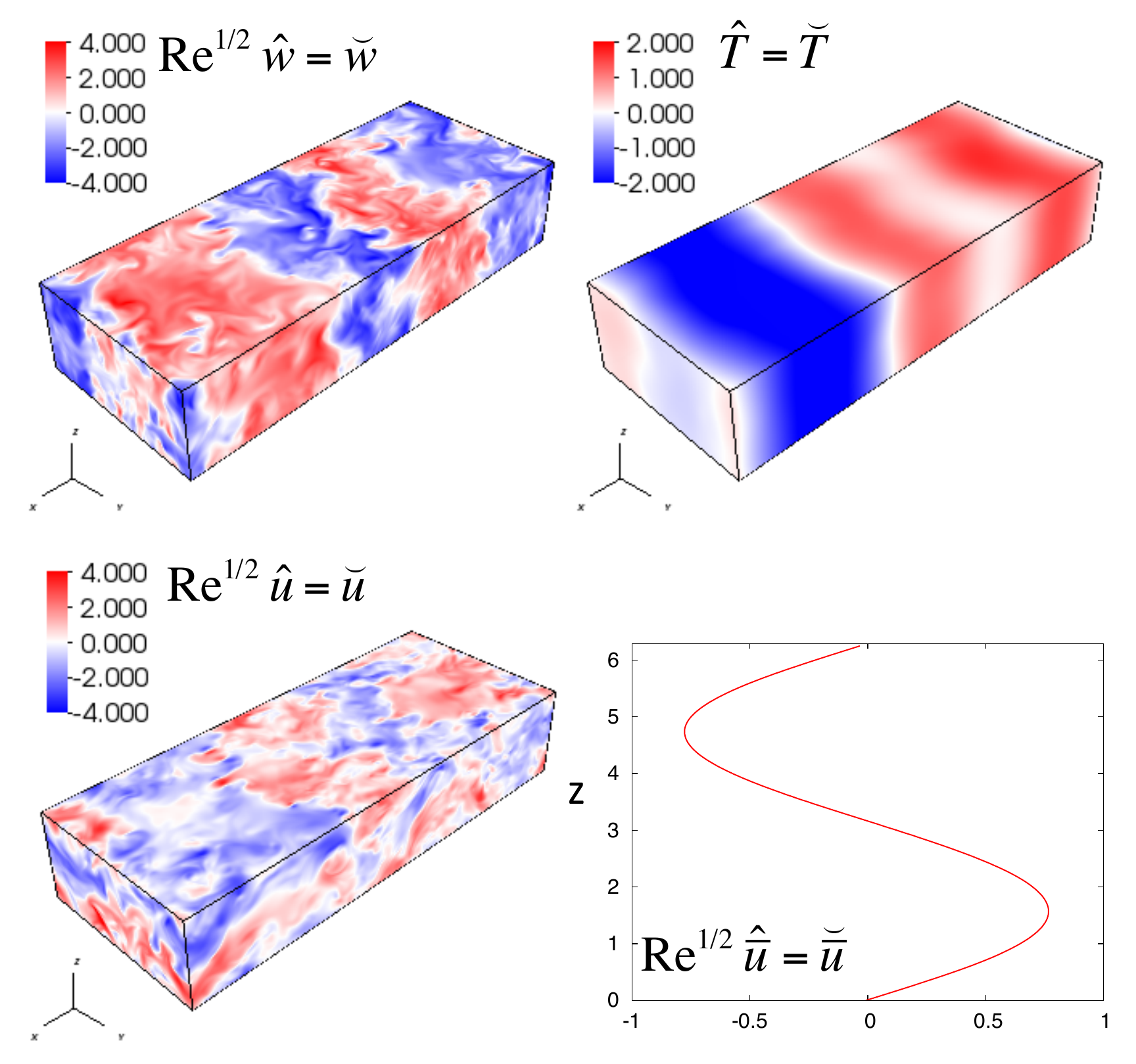}}
  \caption{Simulations snapshots for a run using the standard equations with $\Re = 10^4$, $\Pe = 10$ and $\Ri = 10^{-6}$ (equivalently, $\Re_F = 100$, $\Pe_F = 0.1$ and $\Ri_F = 0.01$, see Section \ref{sec:globalresponse} for detail). From top left to bottom right, we see the vertical velocity fluctuations $\hat w$, the temperature fluctuations $\hat T$, the horizontal velocity fluctuations in the $x$ direction $\hat u$ and the horizontally averaged mean flow in the $x$ direction $\hat{\overline u}$. Note that hatted quantities ($\hat q$) are in the units described in Section \ref{sec:model} while quantities with a breve ($\breve q$) are in the units described in Section \ref{sec:globalresponse}.}
\label{fig:lowRiPelowPe}
\end{figure}

Figure \ref{fig:lowRiPelowPe} shows snapshots of the vertical velocity field $\hat w$, of the horizontal velocity field $\hat u$ and of the temperature perturbations away from the linearly stratified background, $\hat T$, once the forced sheared system has reached a statistically stationary state. In this simulation, $\Pe = 10$ and $\Ri = 10^{-6}$. The very weak stratification, compounded by the relatively important effect of thermal diffusion, implies that the vertical motions proceed un-impeded, and have more-or-less the same scaleheight as that of the imposed forcing (see Section \ref{sec:globalresponse} for a more quantitative estimate of that scaleheight). The scale of the temperature perturbations is commensurate with that of the domain. The horizontal velocity perturbations are significantly stronger than the mean flow, which is hard to recognize in the snapshot. Instead, we see a shear layer that meanders spatially and temporally. The mean velocity, defined as
\begin{equation}
\hat{\overline u}(z) = \frac{1}{t_2 - t_1} \int_{t_1}^{t_2} \left( \frac{1}{L_x L_y} \iint \hat u(x,y,z,t) dxdy \right) dt = \frac{1}{t_2 - t_1} \int_{t_1}^{t_2} \hat{\bar{u}}(z,t) dt \mbox{   ,} 
\label{eq:umean}
\end{equation}
where the time interval $[t_1,t_2]$ over which the data is averaged is taken once the simulation has reached a statistically stationary state, is however well-defined. It is very close to being perfectly sinusoidal, is in phase with the forcing, and its amplitude is of order $\Re^{-1/2}$.

\subsubsection{The stratified limit}

\begin{figure}[h]
  \centerline{\includegraphics[width=0.7\textwidth]{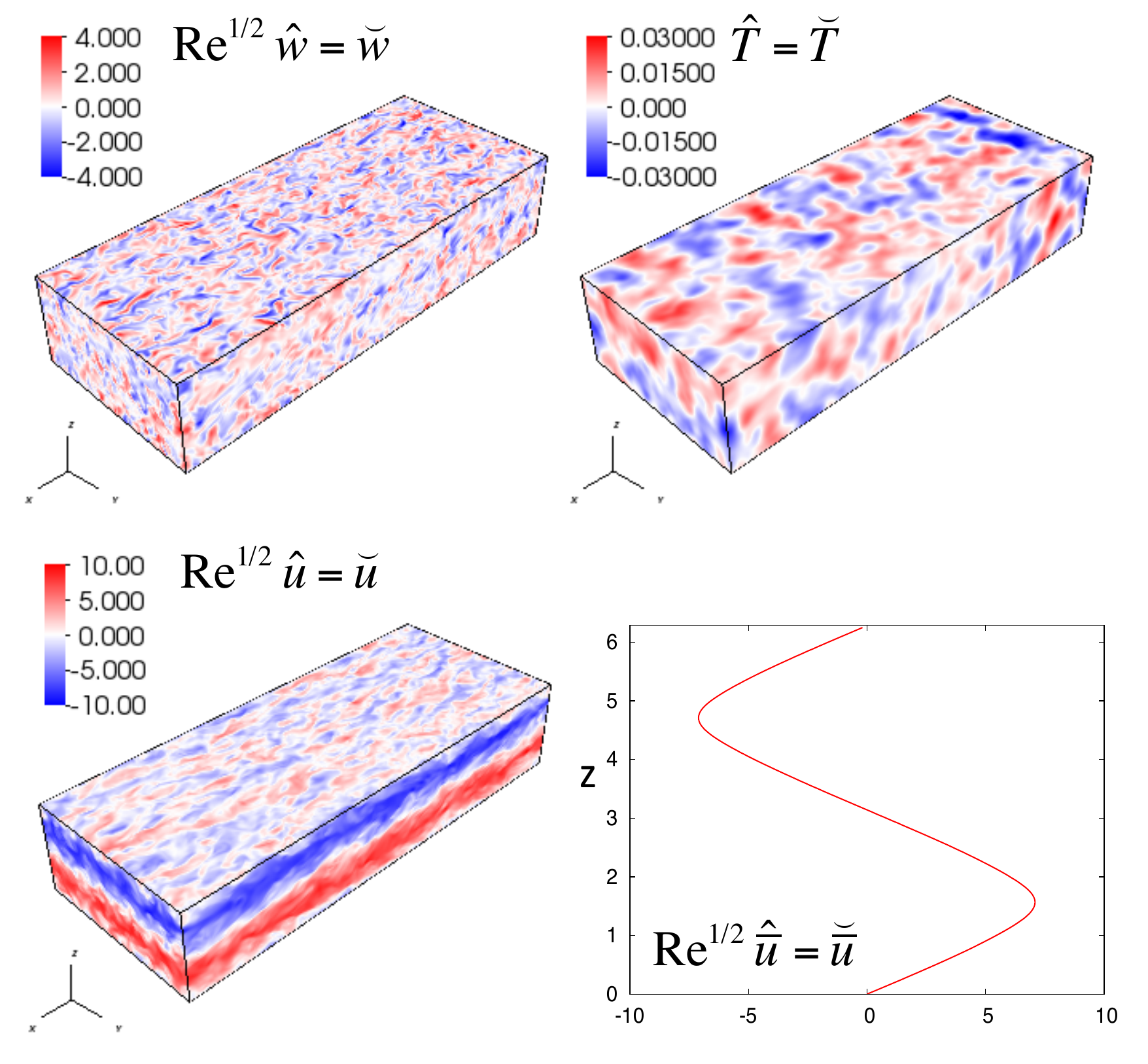}}
  \caption{Simulations snapshots for a run at $\Re = 10^4$, $\Pe = 10$ and $\Ri = 10^{-2}$ (equivalently, $\Re_F = 100$, $\Pe_F = 0.1$ and $\Ri_F = 100$). From top left to bottom right, we see the vertical velocity fluctuations $\hat w$, the temperature fluctuations $\hat T$, the horizontal velocity fluctuations in the $x$ direction $\hat u$ and the horizontally averaged mean flow in the $x$ direction $\hat{\overline u}$. Note that hatted quantities are in the units described in Section \ref{sec:model} while while quantities with a breve ($\breve q$) are in the units described in Section \ref{sec:globalresponse}.}
\label{fig:lowRiPemidPe}
\end{figure}

Figure \ref{fig:lowRiPemidPe} shows snapshots of the same fields as in Figure \ref{fig:lowRiPelowPe} for a simulation with $\Pe = 10$ and $\Ri = 0.01$ so $\Ri\Pe = 0.1$. The results are significantly different from those obtained in the nearly unstratified limit. We see for instance that the scale of the vertical velocity fluctuations is much smaller than before, and the same is true for the temperature fluctuations. The horizontal velocity perturbations are, this time, significantly weaker than the mean flow, and the latter is clearly recognizable in the snapshot. The amplitude of the mean flow is now much larger and its shape is no longer sinusoidal, but instead appears to tend to a piecewise linear profile. It is still essentially in phase with the forcing, however. 

\subsubsection{The strongly stratified limit}
\label{sec:asymm}

\begin{figure}[h]
  \centerline{\includegraphics[width=0.5\textwidth]{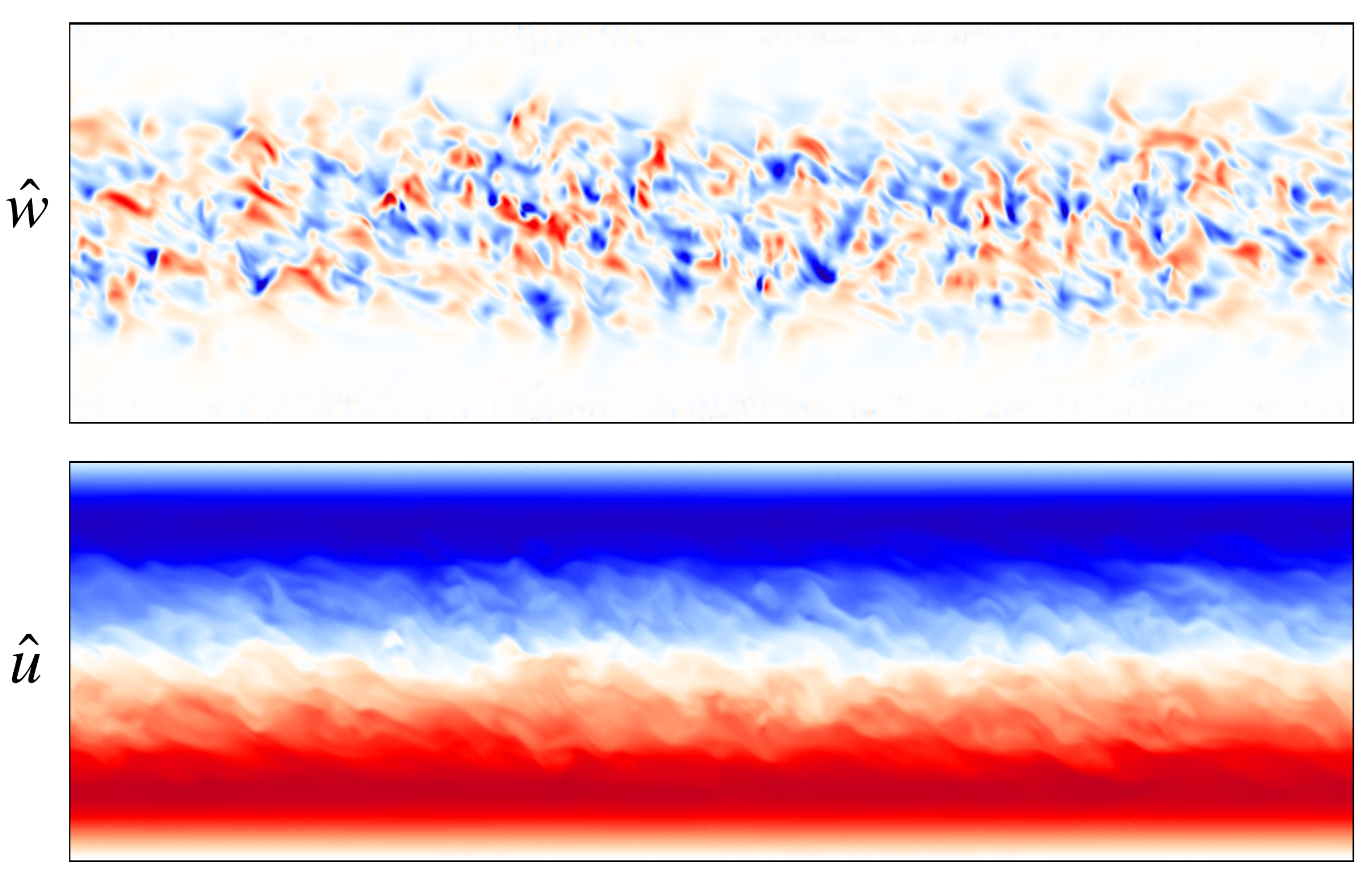}}
  \caption{Snapshots of $\hat u$ and $\hat w$ (as a function of $(x,z)$ in the $y=0$ plane), obtained with LPN equations at $\Re = 10^4$, and $\Ri\Pe = 1.8$ (equivalently, $\Re_F = 100$ and $\Ri_F\Pe_F = 180$). The mean flow in the $x$ direction, for this simulation, is show in Figure \ref{fig:ubarprofiles} for comparison. It is clearly no longer symmetric but becomes skewed. The regions of strongest mean shear are essentially laminar, while the turbulence subsists in the region of weaker shear (near the middle of the box). }
\label{fig:lowRiPehighPe}
\end{figure}

As discussed by \citet{Garaudal15}, turbulent solutions exist for values of $\Ri$ beyond the threshold for linear stability. These solutions can only be obtained by careful continuation of previous solutions, progressively increasing $\Ri$ (equivalently $\Ri\Pe$ in the LPN equations). Using the LPN equations, we have in fact been able to push into the linearly stable regime somewhat further than \citet{Garaudal15} did, and found that, for very strongly stratified flows ($\Ri\Pe > 1.6$ at $\Re = 10^4$), the system dynamics change once again quite dramatically. This is illustrated in Figure \ref{fig:lowRiPehighPe} (see also Figure \ref{fig:ubarprofiles}), which shows that the mean flow no longer has the same symmetries as the imposed force, and adopts instead a new skewed state where the minima and maxima are shifted away from their original positions. This shift effectively enlarges one of the regions of near-constant shear, and creates two thinner (and therefore stronger) shear layers on either side. Surprisingly, we see from the flow snapshots that the stronger/thinner shear layers become laminar, while the turbulence is confined to the weaker/wider one. Why this symmetry breaking occurs, and what stabilizes the region of strongest shear, remain to be determined. %However, this example clearly illustrates that a purely local one-to-one relationship between the shearing rate and the turbulent energy cannot exist (if it did, it would predict some degree of turbulence in the strongly sheared region). 

\subsection{Typical results at high P\'eclet number}
\label{sec:qualreshighPe}

We now look at typical results in the limit of very large P\'eclet numbers, here for $\Pe = 1000$. The unstratified limit in that case is still achieved whenever $\Ri\Pe < 10^{-3}$ (e.g. for $\Ri < 10^{-6}$ if $\Pe = 1000$), and is the same as the one for low P\'eclet numbers described in Section \ref{sec:qualreslowPe}. For larger $\Ri$, however, we observe that instead of settling into a relatively regular statistically stationary state with weak fluctuations around the mean, the system adopts a quasi-periodic behavior that cycles between intense mixing events that destroy the existing shear and render it too weak to maintain turbulence, and fairly quiescent periods during which the system is close to laminar and where the forcing gradually amplifies the shear.  During these laminar periods, the mean flow velocity grows linearly with time. This regime will be discussed in more detail in a forthcoming publication. Indeed, while interesting for many reasons, it has only been observed to exist so far in the case where $\Pe$ is extremely large, and is therefore somewhat outside of the scope of this paper. It is worth noting, however, that similar quasi-periodically bursting stratified shear flows have recently been observed in laboratory experiments by \citet{MeyerLinden14}. 

\begin{figure}[h]
  \centerline{\includegraphics[width=0.35\textwidth]{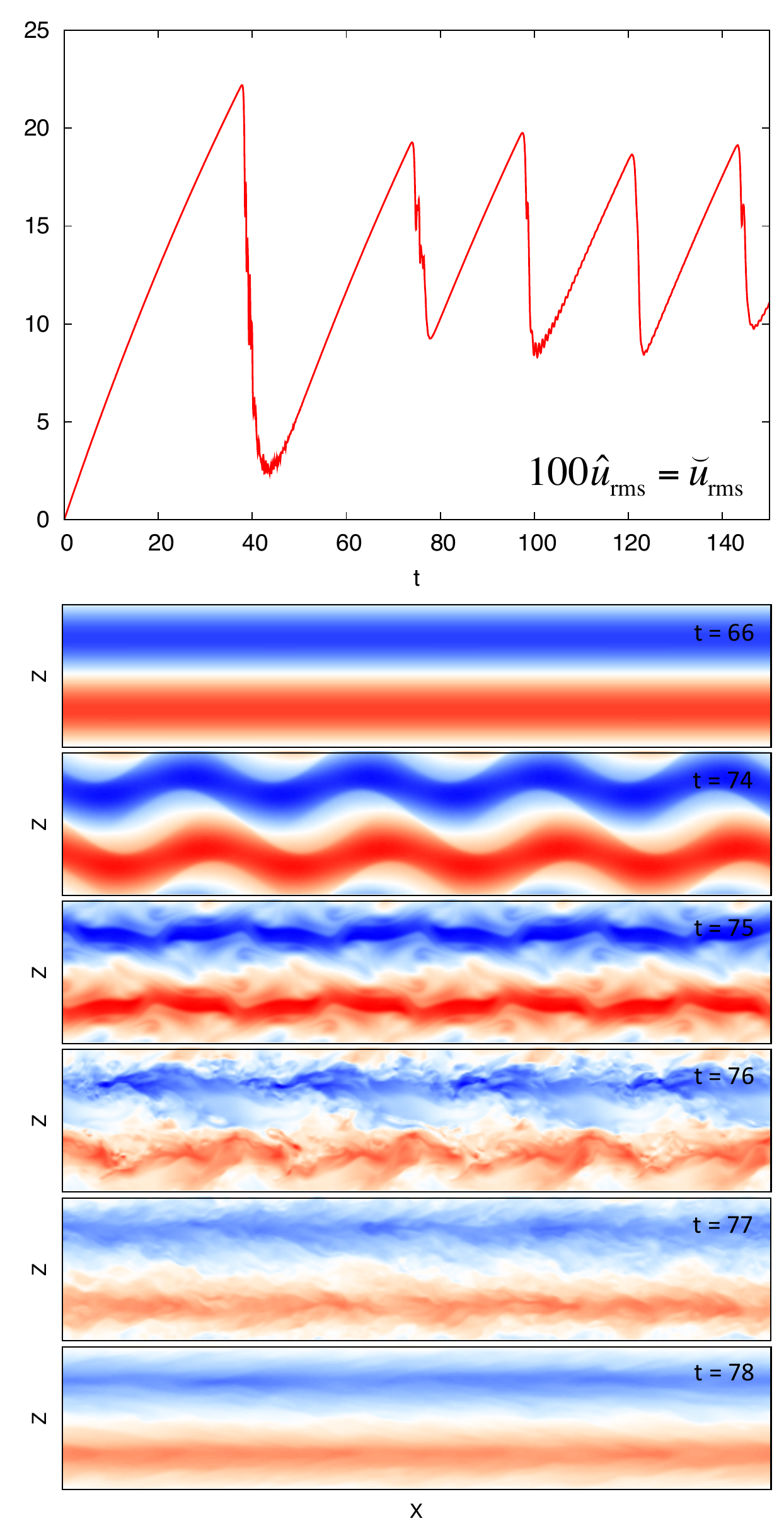}}
  \caption{Temporal evolution of the r.m.s. flow velocity $\hat u_{\rm rms}$ in a simulation with $\Re = 10^4$, $\Pe = 10^3$ and $\Ri = 0.01$ (equivalently, $\Re_F = 100$, $\Pe_F = 10$ and $\Ri_F = 100$), as well as snapshots of the horizontal velocity field $\hat u$ at selected times. This is a typical example of high P\'eclet number bursting behavior. The time intervals where $\hat u_{\rm rms}$ grows linearly correspond to the times where the fluid is essentially laminar, while the sharp drops correspond to brief turbulent mixing events. The snapshot illustrate the dynamics around the second mixing event.}
\label{fig:bursts}
\end{figure}

\section{The validity of the LPN equations}
\label{sec:globalresponse}

We now turn to a more quantitative analysis of our simulations. In this Section, we focus on comparing the results obtained using the LPN equations with those obtained using the standard equations.

\subsection{Comparison between the LPN equations and the full equations}

In order to compare simulations using the LPN equations and those using the standard equations more quantitatively, we now look at typical global properties of the turbulent flow, such as the r.m.s. velocity, and the typical vertical eddy lengthscale in each of the available simulations. 

We define $\hat u_{\rm rms}$ as the non-dimensional instantaneous r.m.s. velocity of the flow, 
\begin{equation}
\hat u_{\rm rms}(t) = \left( \frac{1}{V} \iiint | \hat \bu(\bx,t) |^2 d^3 \bx \right)^{1/2} \equiv \langle | \hat \bu(\bx,t) |^2 \rangle^{1/2} \mbox{  ,}
\label{eq:urmsinst}
\end{equation}
where $V$ is the volume of our computational domain. The $\langle \cdot \rangle$ notation will be used hereafter to denote any volume average. 
We also define $\hat l_v$ as the typical non-dimensional vertical scale of the energy-bearing vertical fluid motions \citep[see for instance][p. 105]{Batchelor53}, with 
\begin{equation}
\hat l_v(t) =  \frac{\sum_{\hat k_x} \sum_{\hat k_y} \sum_{\hat k_z \ne 0} E_{\hat w}(\hat \bk,t) \hat k_z^{-1}  }{\sum_{\hat k_x} \sum_{\hat k_y} \sum_{k_z \ne 0} E_{\hat w}(\hat \bk,t)  }  \mbox{  ,}
\label{eq:lv1}
\end{equation}
where $E_{\hat w}(\hat \bk,t)$ is the instantaneous kinetic energy of the vertical velocity field associated with wavenumber $\hat \bk = (\hat k_x,\hat k_y,\hat k_z)$. Note that there are other more-or-less equivalent ways of defining $\hat l_v$ (see Section \ref{sec:num2} for detail). 

\begin{figure}[h]
  \centerline{\includegraphics[width=0.49\textwidth]{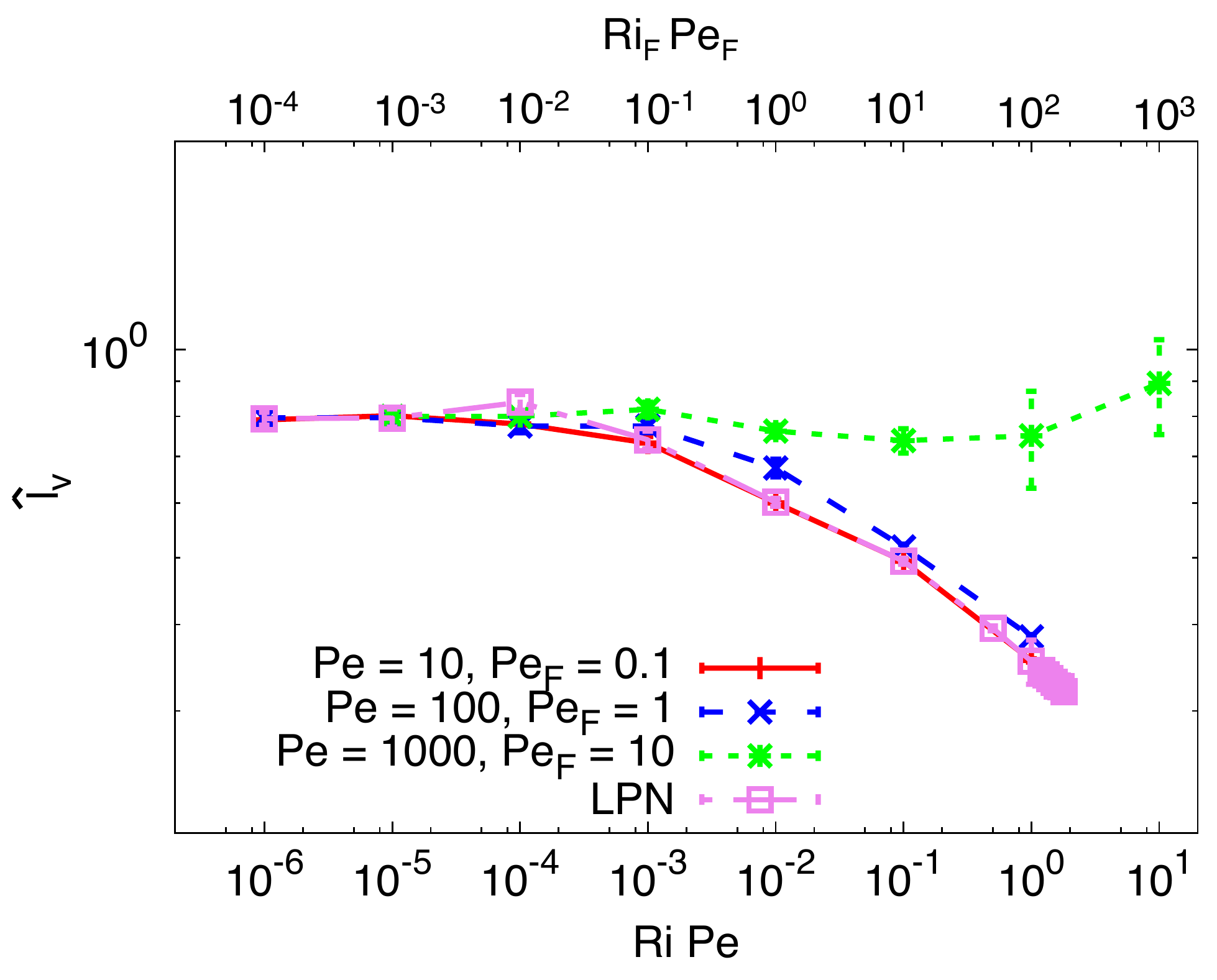}\quad\includegraphics[width=0.5\textwidth]{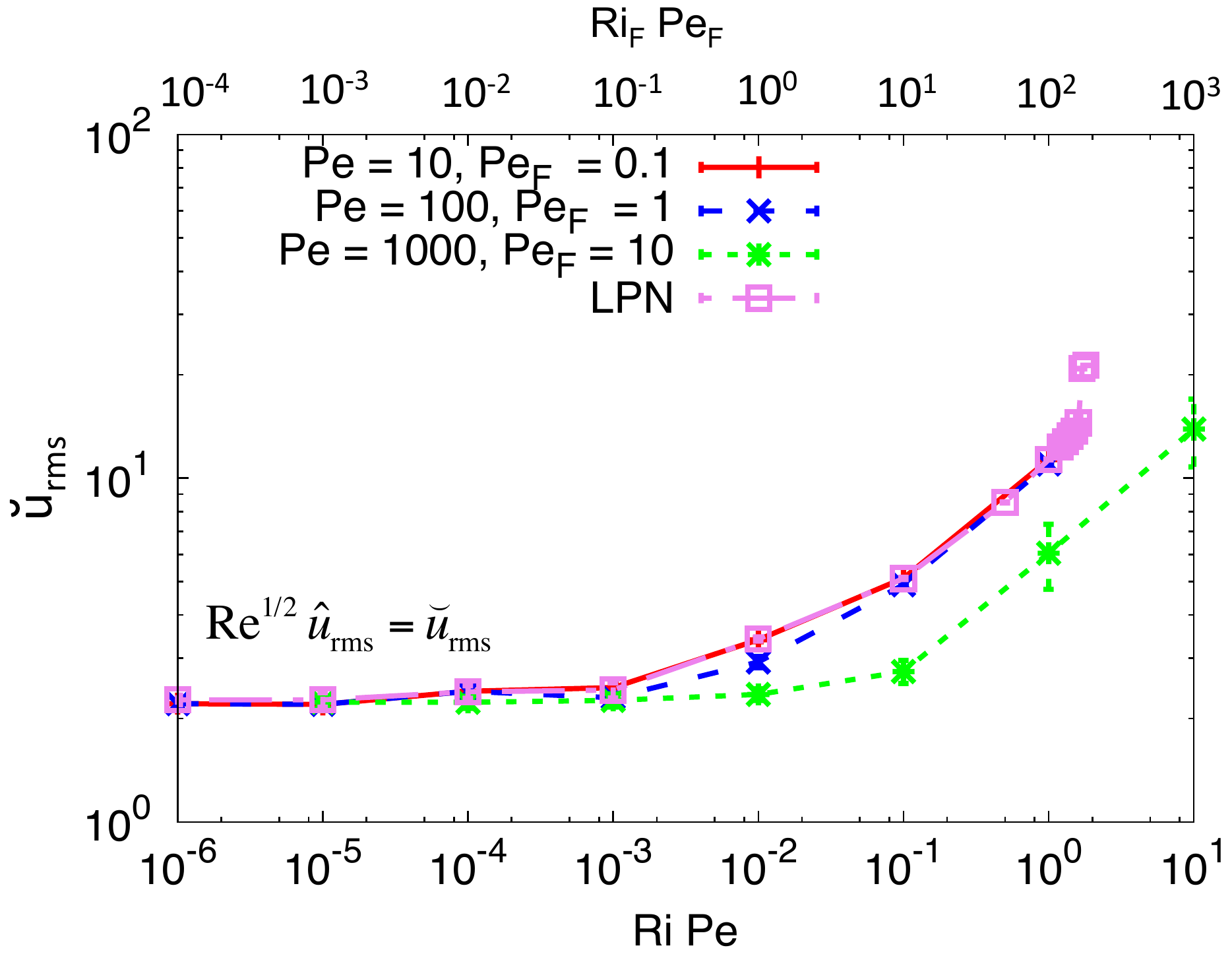}}
  \caption{Mean values of the vertical eddy lengthscale $\hat l_v$ and of the r.m.s. velocity $\hat u_{\rm rms}$ once they system has reached a quasi-steady state, as a function of $\Ri \Pe$ and $\Ri_F \Pe_F$ (see Section \ref{sec:globalresponse}). The error bars represent the r.m.s. fluctuations around the mean value. The LPN equations are a good approximation to the full equations for up to $\Pe = 100$, or equivalently, $\Pe_F = 1$. In the limit where $\Ri_F\Pe_F \ll 1$, we find that $\hat l_v \simeq 0.8\hat k^{-1}$ and $\hat u_{\rm rms} \simeq 0.022$ (equivalently, $\breve u_{\rm rms} = 2.2$).   }
\label{fig:lvurms}
\end{figure}

The quantities $\hat u_{\rm rms}$ and $\hat l_v$ thus defined are functions of time. We then take the mean values of $\hat l_v$ and $\hat u_{\rm rms}$ over a significant time interval after the system has reached a statistically stationary state (as we did in equation (\ref{eq:umean}) for instance). The results are presented in Tables \ref{table1} and \ref{table2a}, and shown in Figure \ref{fig:lvurms}, for the standard equations at various P\'eclet numbers, and for the LPN equations. For ease of comparison between the datasets, we plot $\hat l_v$ and $\hat u_{\rm rms}$ as functions of the Richardson-P\'eclet number. Figure \ref{fig:lvurms} reveals a number of interesting facts. 

First, we see that for each curve (corresponding to each value of $\Pe$ investigated), both $\hat l_v$ and $\hat u_{\rm rms}$ asymptote to the respective constants $\hat l_0 \simeq 0.8$ and $\hat u_0 \simeq 0.022$ in the limit of very low $\Ri\Pe$, and these constants appear to be independent of the P\'eclet number. They simply represent the turbulent properties of an unstratified sinusoidally-forced shear flow at $\Re = 10^4$. Second, we see that the effect of the stratification becomes relevant even for the LPN equations when $\Ri\Pe$ is greater than about $10^{-3}$. As $\Ri\Pe$ increases, the typical vertical eddy scale decreases, and the r.m.s. velocity increases.  The system appears to be stable to finite-amplitude instabilities for $\Ri\Pe$ greater than a few, as discussed by \citet{Garaudal15}.  We shall study the properties of the turbulent solutions in Section \ref{sec:num2} in more detail.

\begin{figure}[h]
  \centerline{\includegraphics[width=0.6\textwidth]{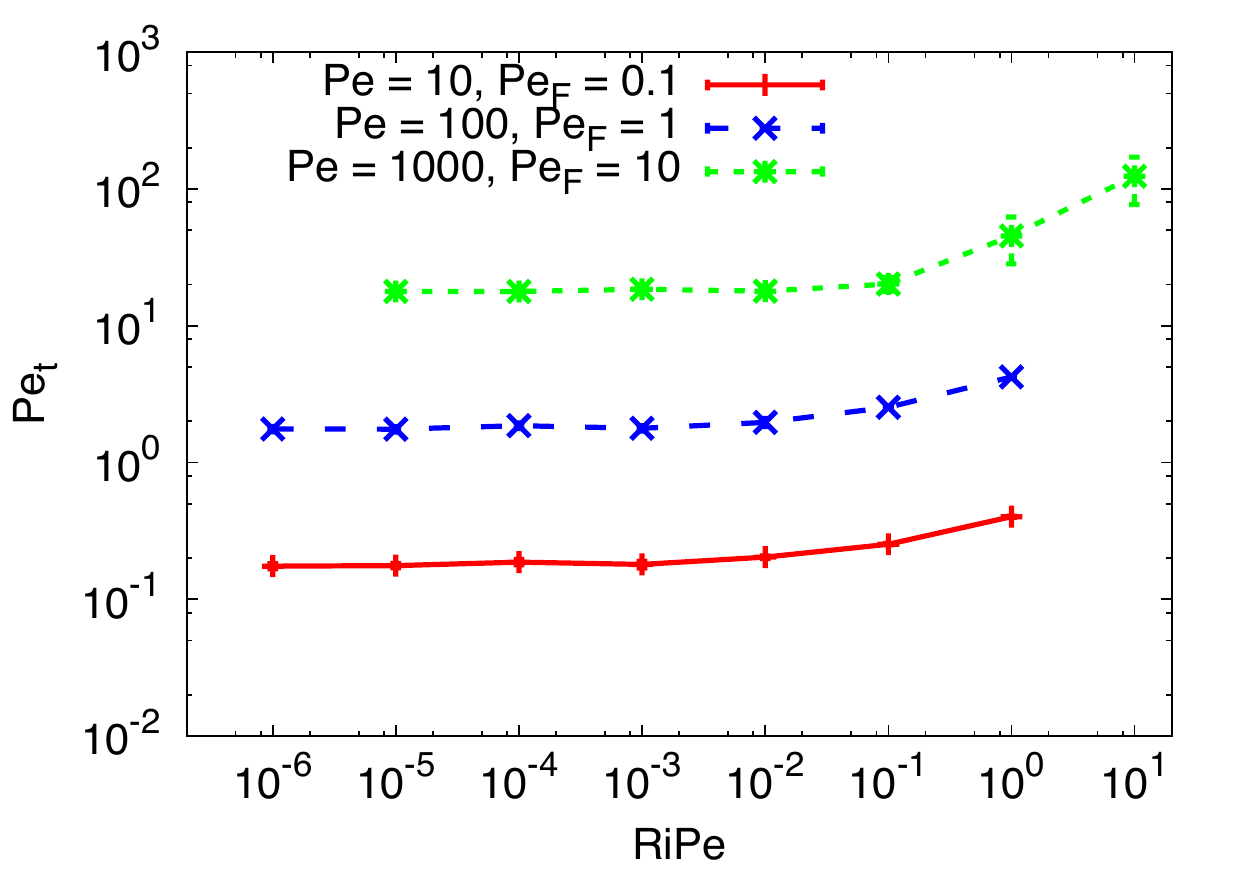}}
  \caption{Turbulent P\'eclet number $\Pe_t = u_{\rm rms} l_v / \kappa_T =   \hat u_{\rm rms} \hat l_v \Pe = \breve{u}_{\rm rms} \breve{l}_v \Pe_F$ as a function of $\Ri\Pe$.}
\label{fig:Pet}
\end{figure}

In both cases (for $\hat l_v$ and for $\hat u_{\rm rms}$), we see quite clearly that the LPN equations are a good-to-excellent approximation of the standard equations up to $\Pe = 100$, but no longer for $\Pe = 1000$. This shows that the domain of validity of the LPN equations extends well beyond $\Pe = 1$. As discussed by \citet{Garaudal15}, this is not entirely surprising. Indeed, $\Pe$ is not a particularly good estimate of the actual turbulent P\'eclet number $\Pe_t = u_{\rm rms} l_v / \kappa_T =  \hat u_{\rm rms} \hat l_v \Pe$ of the unstable flow once a statistically stationary state has been reached. This is demonstrated in Figure \ref{fig:Pet}. We see that $\Pe_t$ is significantly smaller than one for $\Pe =10$, and remains of order one even for $\Pe = 100$. This naturally explains the good correspondence of the LPN simulations and of the standard simulations  up to $\Pe = 100$ seen in Figure \ref{fig:lvurms}, and confirms the argument of  \citet{Lignieres1999}, that the LPN solutions are a good approximation to the true solution provided the {\it turbulent} P\'eclet number $\Pe_t$ is of order one or smaller. 

\subsection{Estimating the turbulent P\'eclet number}
\label{sec:forcingpars}

As discussed in Section \ref{sec:intro}, we would like to estimate, for a given dimensional forcing amplitude $F_0$, shear length scale $k^{-1}$, viscosity $\nu$ and thermal diffusivity $\kappa_T$, what $\Pe_t$ may be. As it turns out, this is actually fairly easy. Indeed, assuming that in the statistically stationary state characteristic of low P\'eclet number simulations there is a balance in the momentum equation between the inertial term and the forcing, such that 
\begin{equation}
\rho_0 (\bu \cdot \bnabla \bu) \cdot \be_x \sim F_0
\end{equation}
in dimensional terms, then we can define a new characteristic dimensional flow amplitude $U_F$ as 
\begin{equation}
U_F = \left(\frac{F_0}{k \rho_0}\right)^{1/2} \mbox{  .}
\end{equation}
Note that $U_F$ no longer depends on any diffusivities, but instead only depends on the characteristics of the forcing (and of the background density). As such, it is a quantity that is more likely to be relevant at high Reynolds numbers than the laminar flow amplitude $U_L$. We can then use $U_F$ to create a new system of units, exactly as in (\ref{eq:unitslaminar}) but with $[u] = U_F$, which then defines new P\'eclet, Reynolds and Richardson numbers as:   
\begin{eqnarray}
\Re_F = \frac{U_F}{k \nu} = \left(\frac{F_0}{k^3 \rho_0 \nu^2} \right)^{1/2}  = \Re^{1/2} \mbox{  ,} \nonumber \\
\Pe_F = \frac{U_F}{k \kappa_T} = \left(\frac{F_0}{k^3 \rho_0 \kappa_T^2} \right)^{1/2} = \Re^{-1/2} \Pe \mbox{  ,}  \nonumber \\
\Ri_F = \frac{N^2}{k^2 U_F^2}  = \frac{N^2 \rho_0}{k F_0 } = \Re \Ri   \mbox{  .}
\label{eq:forcingnumbers}
\end{eqnarray}
In this new system of units, the velocity, pressure and temperature fields are denoted as $\breve{\bu}$, $\breve{p}$ and $\breve{T}$. Note that the unit length and the unit temperature have not changed, so $\hat T = \breve T$ and $\hat l_v = \breve l_v$. Meanwhile, $\breve \bu = \Re_F \hat \bu = \Re^{1/2} \hat \bu$. 

Figure \ref{fig:lvurms} also shows $l_v$ and $u_{\rm rms}$ plotted in this new set of units and against $\Ri_F \Pe_F$ (see the top axis). As we can see, $\breve{u}_{\rm rms}$ is now of order unity for most simulations at low enough $\Ri_F\Pe_F$, and only grows slowly with $\Ri_F \Pe_F$ for $\Ri_F \Pe_F\gg 1$. In other words, $U_F$ seems to be a good predictor for the dimensional r.m.s. velocity of the turbulent shear flow. Furthermore, the apparent transition from weakly stratified to strongly stratified regimes now occurs for $\Ri_F \Pe_F \sim 0.1-1$, which shows that $\Ri_F\Pe_F$ is a more meaningful bifurcation parameter than $\Ri\Pe$ (for which the same transition occurs around 0.01). More importantly, we see in Figure \ref{fig:Pet} that $\Pe_F$ is a good predictor for $\Pe_t$. This is especially true in the weakly stratified limit, where $\Pe_t \sim 1.75 \Pe_F$ but remains also true within an order of magnitude for larger $\Ri_F \Pe_F$. We can now use these results to determine which stars are likely to harbor low P\'eclet number shear layers.

\section{Relevance of diffusive shear instabilities in stars} 
\label{sec:stars}

As found in Section \ref{sec:globalresponse}, in order to determine whether diffusive -- or secular -- shear instabilities could be relevant in stellar interiors, one merely has to calculate $\Pe_F$ using the available information (strength of perturbing force, local diffusivities and local Brunt-V\"ais\"al\"a frequency) and see when the latter is smaller than one. The diffusivities and the local Brunt-V\"ais\"al\"a frequency are known from the background properties of the stellar model. The perturbing force, however, depends on the situation considered -- whether the shear is induced by the differential contraction and expansion of the star,  by tidal forces or other -- and can vary by orders of magnitude accordingly. Bearing this in mind, we can get very rough estimates of the typical values of $\Pe_F$, $\Re_F$ and $\Ri_F$ in the interiors of solar-type stars as follows: 
\begin{eqnarray}
&& \Pe_F= 10^{9}\left( \frac{F_0/\rho_0}{10^{5} \mbox{cm}/ \mbox{s}^2} \right)^{1/2} \left( \frac{k^{-1}}{10^{9} \mbox{cm}} \right)^{3/2} \left( \frac{\kappa_T}{10^7 \mbox{cm}^2/ \mbox{s}} \right)^{-1}  \mbox{  ,} \nonumber \\
&& \Re_F = 10^{15}\left( \frac{F_0/\rho_0}{10^{5} \mbox{cm}/ \mbox{s}^2} \right)^{1/2} \left( \frac{k^{-1}}{10^{9} \mbox{cm}} \right)^{3/2} \left( \frac{\nu}{10 \mbox{cm}^2/ \mbox{s}} \right)^{-1} \mbox{  ,} \nonumber \\
&& \Ri_F = 0.01 \left( \frac{N^2}{10^{-6} \mbox{s}^{-2}} \right) \left( \frac{k^{-1}}{10^{9} \mbox{cm}} \right) \left( \frac{F_0/\rho_0}{10^{5} \mbox{cm}/ \mbox{s}^2} \right)^{-1} \mbox{  ,} 
\end{eqnarray} 
where the numerical values chosen for comparison for $k^{-1}$, $\kappa_T$ and $\nu$ are order-of-magnitude estimates of the solar tachocline properties \citep{Gough07}, a well-known shear layer at the interface between the convection zone and the radiative zone of the Sun \citep{Tachocline2007}. Meanwhile $F_0/\rho_0$ is compared with its local gravity $g \simeq 10^{5}$cm/s$^2$. 

This clearly shows that, unless $F_0/\rho_0$ is many orders of magnitude smaller than $g$, a tachocline-like shear cannot be viewed as a low P\'eclet number flow. And, supposing that $F_0/\rho_0$ were to be of the order of $10^{-13}$, then $\Ri_F$ would be of the order of $10^{20}$. This shows that diffusive shear instabilities are unlikely to play a role in the dynamics of the solar tachocline, and that the latter is most probably stable to its vertical shear\footnote{On the other hand it may still be unstable to its horizontal shear \citep{Watson81,Garaud01}.}.
Another way to have a low $\Pe_F$ would be to consider the possibility of very thin shear layers. Using again the tachocline value of $\kappa_T \sim 10^7$cm$^2$/s, and taking $F_0/\rho_0 \sim 0.1$cm/s$^2$, would require the shear layer to be no thicker than $k^{-1} \sim 10^6$cm (10 km) to be diffusive. Either way, it appears that diffusive shear instabilities are unlikely in the interior of solar-type stars. 

\begin{figure}[h]
  \centerline{\includegraphics[width=0.7\textwidth]{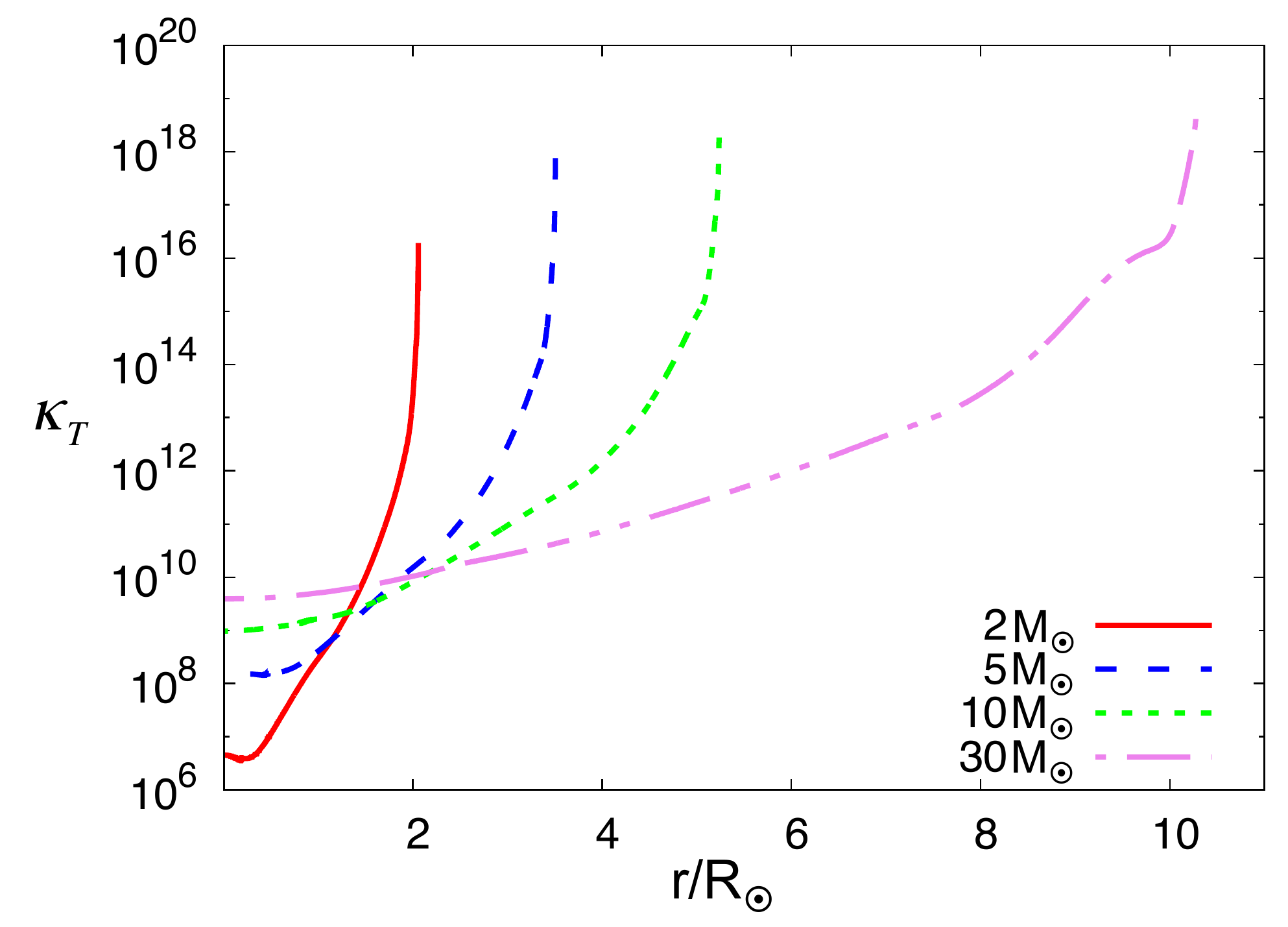}}
  \caption{Thermal diffusivity $\kappa_T$ (in cm/s$^2$) as a function of radius for stars of various masses $M_\star$ ranging from 2$M_\odot$ to $30M_\odot$. }
\label{fig:kappaT}
\end{figure}

In much more massive stars, however, the thermal diffusivity can be much larger. This is illustrated in Figure \ref{fig:kappaT}, which shows $\kappa_T$ as a function of radius for a range of stars on the Main Sequence. Here, $\kappa_T$ is calculated as in \citet{Garaudal15b}. We see that for stars larger than about 10$M_\odot$, an increasingly large fraction of their outer layers has very high thermal diffusivities, with values of $10^{14}$cm$^2$/s or larger. The reason why $\kappa_T$ increases with $M_\star$ is mostly due to the fact that higher-mass stars typically have lower densities and higher temperatures than lower-mass stars at the same fractional mass coordinate. 

With the same scalings as before, but with $\kappa_T$ compared with $10^{14}$cm$^2$/s instead, we get
\begin{equation}
\Pe_F= 100 \left( \frac{F_0/\rho_0}{10^{5} \mbox{cm}/ \mbox{s}^2} \right)^{1/2} \left( \frac{k^{-1}}{10^{9} \mbox{cm}} \right)^{3/2} \left( \frac{\kappa_T}{10^{14} \mbox{cm}^2/ \mbox{s}} \right)^{-1}  \mbox{  ,}
\end{equation} 
This time, $\Pe_F$ values smaller than one can be achieved with a reasonably thin shear layer ($k^{-1} \sim 10^7$cm), or with a reasonable force of $F_0 / \rho_0 \sim 10$cm/s$^2$. This implies that diffusive shear instabilities could be relevant in the outer layers of high-mass stars. At the same time, the values of $N^2$ in the radiative layers of high-mass stars range between $10^{-8}$ and $10^{-7}$, so the bifurcation parameter
\begin{equation}
 \Ri_F \Pe_F = 0.1 \left( \frac{N^2}{10^{-7} \mbox{s}^{-2}} \right) \left( \frac{k^{-1}}{10^{9} \mbox{cm}} \right)^{5/2} \left( \frac{F_0/\rho_0}{10^{5} \mbox{cm}/ \mbox{s}^2} \right)^{-1/2}   \left( \frac{\kappa_T}{10^{14} \mbox{cm}^2/ \mbox{s}} \right)^{-1} \mbox{  ,} 
\end{equation} 
will likely be in the range $10^{-3} -10^{3}$ depending on the exact value of $\kappa_T$, $N^2$, $F_0$ and $k$ used. In other words, $\Ri_F \Pe_F$ could in principle span most of parameter space between the nearly unstratified limit and the strongly stratified limit.  With this in mind, we now go back to the simulations and look at their basic transport properties in more detail.

%The unit velocity selected here naturally occurs when there is a balance between the amplitude of the inertial term . The system of equations then becomes 
%\begin{eqnarray}
%&& \frac{\p \bu}{\p t} +  \bu \bcdot \bnabla \bu = -  \bnabla  p + \Ri_f \, T \be_z + \frac{1}{\Re_f} \nabla^2 \bu- \sin(z) \be_x   \mbox{  ,} \label{eq:originallaminar1}\\
%&& \bnabla \bcdot  \bu = 0 \mbox{  ,}\\
%&& \frac{\p T}{\p t} + \bu \cdot \bnabla T + w = \frac{1}{\Pe_f} \nabla^2 T\mbox{  ,}
%\label{eq:originalforcing}
%\end{eqnarray}
%where the non-dimensional numbers $\Pe_f$, $\Re_f$ and $\Ri_f$ are 
%\begin{equation} 
%\Ri_f = \alpha g |T_{0z}-T_{0z}^{\rm ad}| k^{-1} \frac{\rho_0}{ F_0} = \frac{N^2}{k^2 [u]^2} , \, \Re_f =  \left(\frac{F_0 }{\nu^2 k^3 \rho_0}\right)^{1/2} = \frac{[u]}{k \nu}, \, \quad \,   \Pe_f =  \left(\frac{F_0 }{\kappa_T^2 k^3 \rho_0}\right)^{1/2}  = \frac{[u]}{k\kappa_T} \mbox{  .}
%\end{equation}
%These numbers are related to the Richardson, Reynolds and P\'eclet number of the flow in the limit where the eddy scale is of the order of forcing scale $k^{-1}$, and where the r.m.s. velocity is of the order of $[u]$. As we shall see in Section 3, this is precisely the case for relatively weakly stratified flows. 

\section{Global properties of the turbulent LPN solutions at low and high $\Ri_F \Pe_F$}
\label{sec:num2}

We now restrict our analysis to flows that have a low predicted P\'eclet number (with $\Pe_F \le 1$), focussing on the results obtained using the LPN equations unless otherwise specified. In all that follows, we adopt the non-dimensionalization based on the forcing, where the velocity is expressed in units of $U_F$, and use $\Ri_F \Pe_F$ and $\Re_F$ as our basic input parameters.  

\subsection{Properties of the turbulent flow}
\label{sec:KErms}

\begin{figure}[h]
\includegraphics[width=\textwidth]{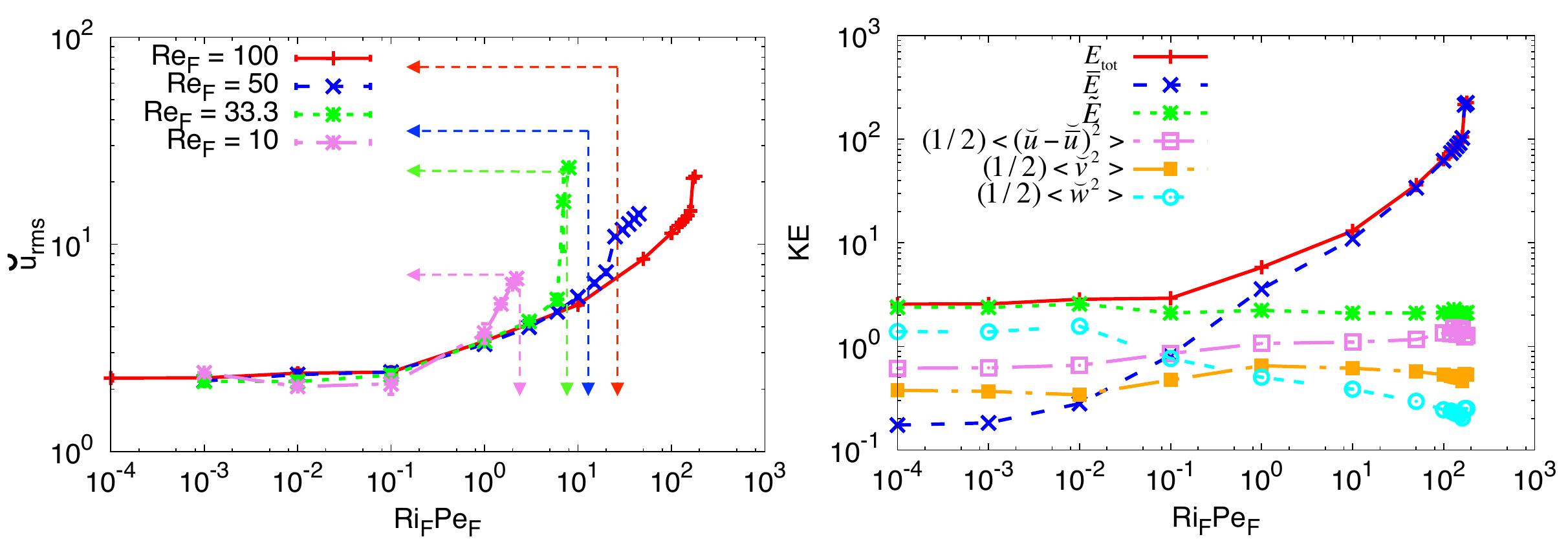} 
  \caption{Left: r.m.s. velocity of the flow for the LPN equations for various $\Re_F$. For each value of $\Re_F$, the vertical arrow marks the limit for linear stability of the flow, while the horizontal arrow marks the value of $\breve u_{\rm rms}$ corresponding to the laminar solution. Note how the lower $\Re_F$ simulations smoothly tend to the laminar solution while the higher $\Re_F$ simulations follow a nonlinearly unstable turbulent branch that ignores the laminar solution entirely. Right: Partition between the different contributions to the total kinetic energy of the flow, for $\Re_F = 100$ and various $\Ri_F \Pe_F$, in the LPN equations. In both figures, the quantities plotted are in the units based on $U_F$. }
\label{fig:urms2}
\end{figure}

Figure \ref{fig:urms2}a shows the r.m.s. velocity of the flow as a function of $\Ri_F\Pe_F$ for various values of $\Re_F$. As already discussed in Section \ref{sec:num}, we find that the r.m.s. velocity becomes independent of $\Ri_F\Pe_F$ in the limit of very low stratification ($\Ri_F\Pe_F \ll 1$). This is expected since the buoyancy force plays an insignificant role in the momentum equation in this limit, so the characteristics of the system should only depend on the remaining parameter $\Re_F$. Interestingly, we find that viscosity is already essentially negligible for $\Re_F \ge 33.3$, with $\breve u_{\rm rms}(\Ri_F\Pe_F \rightarrow 0,\Re_F) \simeq 2.2$. Dimensionally, this implies that $u_{\rm rms} \simeq 2.2 (F_0 / k \rho_0)^{1/2}$ in high $\Re_F$, low $\Ri_F \Pe_F$ shear flows.

Figure \ref{fig:urms2}a also shows that the r.m.s. velocity increases monotonically (albeit not always smoothly, see below) with $\Ri_F \Pe_F$ for all $\Re_F \ge 33.3$. This trend can easily be explained: as the stratification increases, it becomes gradually more difficult to destabilize the forced horizontal flow, so the shear must grow to larger amplitudes before turbulence can set in. As a result, the total kinetic energy in the fluid in the turbulent state also increases. However, it is interesting to note this increase is solely due to the increase in the amplitude of the mean flow. 
%We  also find that the  %as  $u_{\rm rms} \propto (\Ri_f\Pe_f)^{1/3}$. 
%Note again that this scaling is also very tentative, as it is only based on few data points and a very limited range of $\Ri_f \Pe_f$. However, as for $l_v$, we present a plausible theoretical explanation for it in Section \ref{sec:modelexplained}.
Figure \ref{fig:urms2}b demonstrates this by comparing the mean flow kinetic energy, given by the average of 
\begin{equation}
\bar E(t)  =  \frac{1}{L_z} \int \frac{1}{2} \breve{\overline{u}}^2(z,t) dz
\end{equation}
over a suitable time span, with that of entire flow field $\breve \bu$, given by the time average of
\begin{equation} 
E_{\rm tot}(t)  =  \frac{1}{2} \langle|  \breve{\bu}(x,y,z,t) |^2 \rangle  \mbox{  ,}
\end{equation}
and that of the perturbations, given by the time average of
\begin{equation}
\tilde{E}(t)  = E_{\rm tot}(t) - \bar E(t)  \mbox{   .} 
\end{equation}
In the limit $\Ri_F\Pe_F \ll 1$, we find that the kinetic energy in the mean flow is only about 1/16 of the total kinetic energy. In this limit, most of the energy is in the perturbations. By contrast,  when $\Ri_F\Pe_F \gg 1$ the mean flow is the major contributor to the total kinetic energy in the system. 
The properties of the mean flow are discussed in more detail in Section \ref{sec:mean}.

Going back to Figure \ref{fig:urms2}a, we see that the manner in which $\breve u_{\rm rms}$ increases with $\Ri_F\Pe_F$ is notably different between the low Reynolds number runs ($\Re_F = 10$ and $\Re_F = 33.3$) and the high Reynolds number runs ($\Re_F = 50$ and $\Re_F = 100$). 
For $\Re_F = 10$ and $\Re_F = 33.3$, $\breve u_{\rm rms}$ is a smooth function of $\Ri_F\Pe_F$, and the respective curves gradually approach the point of marginal linear stability (which is marked by the intersection of the horizontal and vertical arrows). In other words, the turbulence gradually dies away as the system approaches that marginal state. 

For $\Re_F = 50$ and $\Re_F = 100$, however, the linear stability threshold is no longer relevant, and turbulent solutions extend significantly into the linearly stable region of parameter space \citep{Garaudal15}. For both sets of simulations, we see that shortly after crossing the threshold for marginal linear stability, the $\breve u_{\rm rms}(\Ri_F \Pe_F)$ curve has a very sharp step, followed by another gradual incline. This step corresponds to the transition discussed in Section \ref{sec:asymm} from a symmetric to a skewed flow profile, the latter having a significantly higher total kinetic energy. The shape of the step being somewhat reminiscent of half a hysteresis curve, we have tried to find evidence of multiple equilibria in this system. However, we have not found any: the functional dependence of $\breve u_{\rm rms}$ on $\Ri_F\Pe_F$ is the same whether $\Ri_F\Pe_F$ is gradually increased or gradually decreased across the step\footnote{It remains possible that there are indeed multiple equilibria, but if they exist, they only do so over a range of $\Ri_F \Pe_F$ too narrow for us to find.}.  Finally, Figure \ref{fig:urms2}a shows that the position of the step depends on $\Re_F$, showing that viscosity plays a role in this transition. This is not a surprising result, given that the transport of momentum across the strong/laminar portion of the shear flow (see Section \ref{sec:asymm}) has to be diffusive. 

Figure \ref{fig:urms2}b studies the various contributions to the total kinetic energy in the system, for the runs with $\Re_F = 100$. We find that even though the total kinetic energy increases with $\Ri_F\Pe_F$, the kinetic energy of the perturbations $\tilde{E}$ remains more-or-less constant for all simulations. This is a rather remarkable result (since $\Ri_F \Pe_F$ varies by more than six orders of magnitude) which appears to be one of the defining properties of forced shear flows at low P\'eclet numbers. A closer inspection of the contributions to the turbulent kinetic energy from both streamwise horizontal motions, cross-stream horizontal motions and vertical motions, shows that even though their total remains constant, the kinetic energy in the vertical motions drops slowly with increasing $\Ri_F\Pe_F$, while that in the horizontal motions increases slightly to compensate. What controls the partitioning of the turbulent kinetic energy, which is always of order one in this system, into vertical and horizontal perturbations respectively, remains to be determined. 

 \subsection{Mean flow properties}
\label{sec:mean}

\begin{figure}[h]
  \centerline{\includegraphics[width=0.6\textwidth]{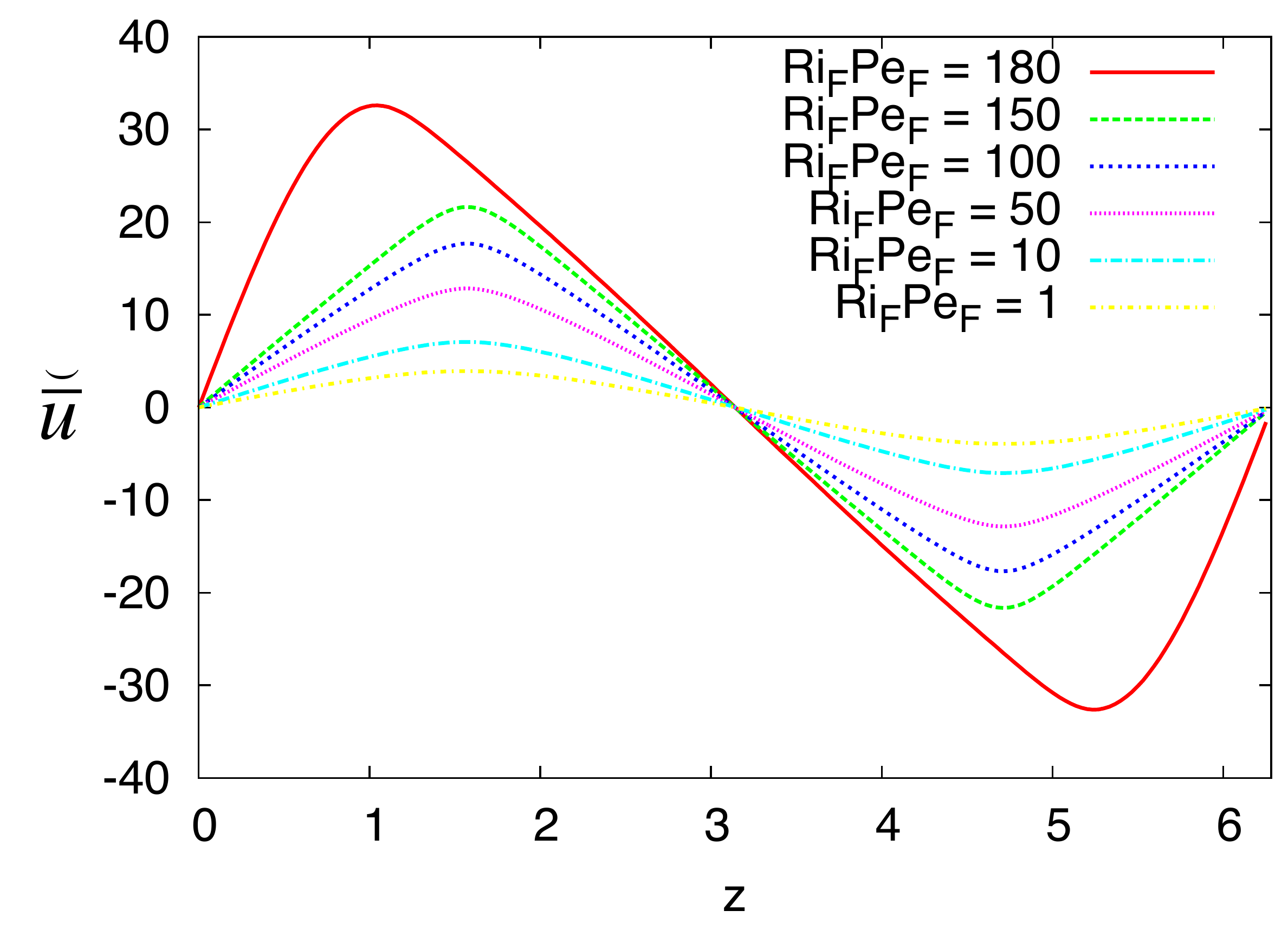}}
    \caption{Variation of the mean flow profile and amplitude as a function of $\Ri_F\Pe_F$ for $\Re_F = 100$, using the LPN equations. }
\label{fig:ubarprofiles}
\end{figure}

As discussed in Section \ref{sec:num}, for fixed Reynolds number, both the amplitude and shape of the mean flow change significantly as $\Ri_F \Pe_F$ increases from the unstratified limit up to the critical value above which turbulent solutions cease to exist. This is illustrated in Figure \ref{fig:ubarprofiles}. For low $\Ri_F \Pe_F$, the profiles are very close to being perfectly sinusoidal and are in phase with the forcing. As $\Ri_F\Pe_F$ increases beyond one, the profiles become more and more  triangular, but the maxima and minima of $ \breve{\overline{u}}$ remain aligned with the maxima and minima of the imposed forcing (i.e. at $z = \pi/2$ and $z = 3\pi/2$). Finally, for $\Ri_F \Pe_F = 170$ (not shown) and $\Ri_F \Pe_F = 180$ (shown), the profiles become asymmetric, as discussed in Section \ref{sec:asymm}. 

\begin{figure}[h]
  \centerline{\includegraphics[width=0.6\textwidth]{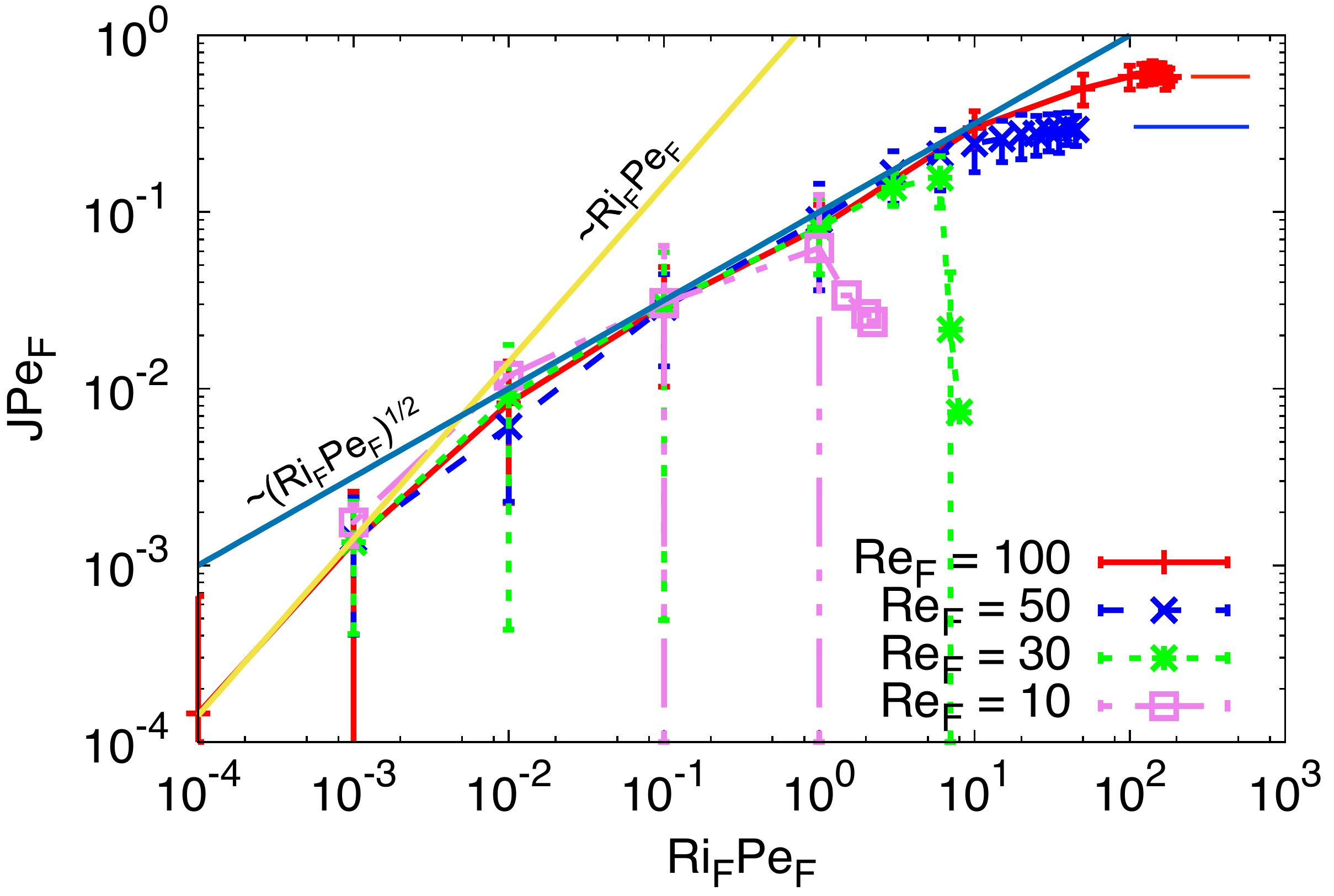}}
    \caption{Variation of $J \Pe_F$ as a function of $\Ri_F\Pe_F$. The horizontal lines mark the apparent saturation values for each curve in the nonlinearly unstable regime ($\Re_F  \ge 50$ and large $\Ri_F \Pe_F$), and are consistent with $J \Pe_F \simeq 0.006 \Re_F$. The line fitting the data at very low $\Ri_F \Pe_F$ has $J \Pe_F = 1.4 \Ri_F \Pe_F$, the line fitting the data at intermediate values of $\Ri_F \Pe_F$ has $J \Pe_F = 0.1 (\Ri_F \Pe_F)^{1/2}$. }
\label{fig:JPe}
\end{figure}

Interestingly, the shear at $z = \pi$ remains a smooth function of $\Ri_F \Pe_F$ despite the dramatic change of behavior associated with the loss of symmetry. This can be seen more clearly in Figure \ref{fig:JPe}, which shows the product of $\Pe_F$ with the gradient Richardson number $J$ as a function of $\Ri_F \Pe_F$. $J$ is calculated from the amplitude of the mean shear at $z = \pi$, as
\begin{equation}
J = \frac{\Ri_F}{\breve S^2} \mbox{   where, here,  } \breve S = \left|\frac{d \breve{\overline{u}}}{d z}\right|_{z = \pi} \mbox{   .} 
\label{eq:gradRich}
\end{equation}
We see that for the larger $\Re_F$ runs (which are the only ones we believe to be relevant for stellar interiors), $J \Pe_F$ grows smoothly as $\Ri_F \Pe_F$ increases. For very low $\Ri_F\Pe_F$, $J \Pe_F$ is proportional to $\Ri_F \Pe_F$, which can easily be explained from the fact that the mean flow amplitude, and therefore the mean shear, are independent of $\Ri_F \Pe_F$ in that limit (see Figure  \ref{fig:lvurms}b). For intermediate values of $\Ri_F \Pe_F$, $J \Pe_F$ seems to scale like $(\Ri_F \Pe_F)^{1/2}$, a result which remains to be explained. 

For large $\Ri_F \Pe_F$, $J \Pe_F$ appears to saturate at a value that is proportional to $\Re_F$. This result could be consistent with the notion that a strongly stratified shear flow adapts itself to satisfy a theoretical marginal stability\footnote{It is crucial to note, however, that this has nothing to do with marginal stability to linear perturbations, since the system continues to be turbulent well-into the region of linear stability.} criterion of the kind $J \Pr  = (J \Pr)_c$ where $(J \Pr)_c$ is independent of $\Re_F$, as in (\ref{eq:jzahn}) for instance. The two data points that are available (for $\Re_F = 50$ and 100) suggest that for large enough stratification, the flow satisfies
\begin{equation}
J \Pr \simeq (J \Pr)_c \simeq 0.006 \mbox{   .}
\label{eq:numstabcrit}
\end{equation}
This is in fact remarkably similar to the criterion proposed by Zahn from heuristic energy stability arguments (see equation \ref{eq:jzahn}). It is unfortunate, however, that we have so few data points in the strongly stratified limit, which prevents us from firmly establishing the validity of (\ref{eq:numstabcrit}). The forced shear layer rapidly becomes fully stable for larger values of $\Ri_F \Pe_F$ (at least in the case of this sinusoidal forcing, see Section \ref{sec:ccl} for detail), so we are unable to determine whether (\ref{eq:numstabcrit}) continues to hold for even stronger levels of stratification. Numerical simulations at larger $\Re_F$ may help resolve this problem, since they are likely to exhibit instability for larger values of $\Ri_F \Pe_F$. However, they are computationally prohibitive to date. 

Finally, we note that while the transition from $J\Pe_F \sim \Ri_F \Pe_F$ to $J\Pe_F \sim (\Ri_F \Pe_F)^{1/2}$ is independent of viscosity and always occurs around $\Ri_F \Pe_F \sim 0.01$, the transition from $J\Pe_F \sim (\Ri_F \Pe_F)^{1/2}$  to $J \Pe_F \sim (J \Pe_F)_c$ occurs at progressively larger values of $\Ri_F \Pe_F$ as $\Re_F$ increases. For this particular problem, we find that, {\it very roughly}, 
\begin{eqnarray}
J\Pe_F \simeq 1.4 \Ri_F \Pe_F  \mbox{  for }  \Ri_F \Pe_F \le 0.005 \mbox{   ,} \nonumber \\
J\Pe_F \simeq 0.1 (\Ri_F \Pe_F)^{1/2}  \mbox{  for }  0.005 \le \Ri_F \Pe_F \le (0.06 \Re_F)^2 \mbox{   ,}  \nonumber \\
J\Pe_F \simeq 0.006 \Re_F   \mbox{  for  }  (0.06 \Re_F)^2 \le \Ri_F \Pe_F  \le (\Ri_F \Pe_F)_c \sim \Re_F  \mbox{   ,}
\label{eq:regimes}
\end{eqnarray}
where the numerical constants were determined by fitting the data, and the ranges of validity were determined by requiring approximate continuity of $J \Pe_F$ with $\Ri_F \Pe_F$. 
The upper limit for the existence of turbulent solutions, $(\Ri_F \Pe_F)_c$, was determined numerically to be a constant of order unity times $\Re_F$, for the sinusoidal forcing selected \citep{Garaudal15}. Taken at face value, equation (\ref{eq:regimes}) would imply that the third regime, where $J\Pe_F \simeq 0.006 \Re_F$, may disappear altogether for large enough $\Re_F$. However, given our current lack of explanation for the scalings observed in the intermediate regime, and the uncertainties in the determination of $(\Ri_F \Pe_F)_c$ \citep[see][for detail]{Garaudal15}, we caution the reader against using the formulas given in equation (\ref{eq:regimes}) too far outside of the range of the available data. Within that range, however, (\ref{eq:regimes}) can be used to predict $J$ (and hence the mean shear) resulting from a given forcing, and a given set of fluid parameters (viscosity, thermal diffusivity and stratification).  

\subsection{Typical scale and shape of the turbulent eddies}

We now study in more detail the scale and shape of the turbulent eddies and how they vary with input parameters. 
In Section \ref{sec:globalresponse}, we defined the vertical scale $\hat l_v$ (or equivalently $\breve l_v$ since the two are the same) of the turbulent eddies, and showed their variation with $\Ri_F\Pe_F$ in the LPN simulations in Figure \ref{fig:lvurms}. By analogy, we can also define the typical horizontal scales $\breve l_x$ and $\breve l_y$ as 
\begin{eqnarray}
\breve l_x =  \frac{\sum_{\breve k_x\ne 0} \sum_{\breve k_y} \sum_{\breve k_z} \breve E_u(\breve \bk) \breve k_x^{-1}  }{\sum_{\breve k_x\ne 0} \sum_{\breve k_y} \sum_{\breve k_z}  \breve E_u(\breve \bk)}  \mbox{  ,} \nonumber \\
\breve l_y =  \frac{\sum_{\breve k_x} \sum_{\breve k_y\ne 0} \sum_{\breve k_z} \breve E_v(\breve \bk)\breve  k_y^{-1}   }{\sum_{\breve k_x} \sum_{\breve k_y\ne 0} \sum_{\breve k_z}  \breve E_v(\breve \bk)  }  \mbox{  ,} 
\end{eqnarray} % Need to decide on notation. 
where $\breve E_u(\breve \bk)$ and $\breve E_v(\breve \bk)$ are the kinetic energies associated with fluid motion in the $x-$ and $y-$directions respectively, with wavenumber $\breve \bk$ . 

\begin{figure}[h]
  \centerline{\includegraphics[width=0.6\textwidth]{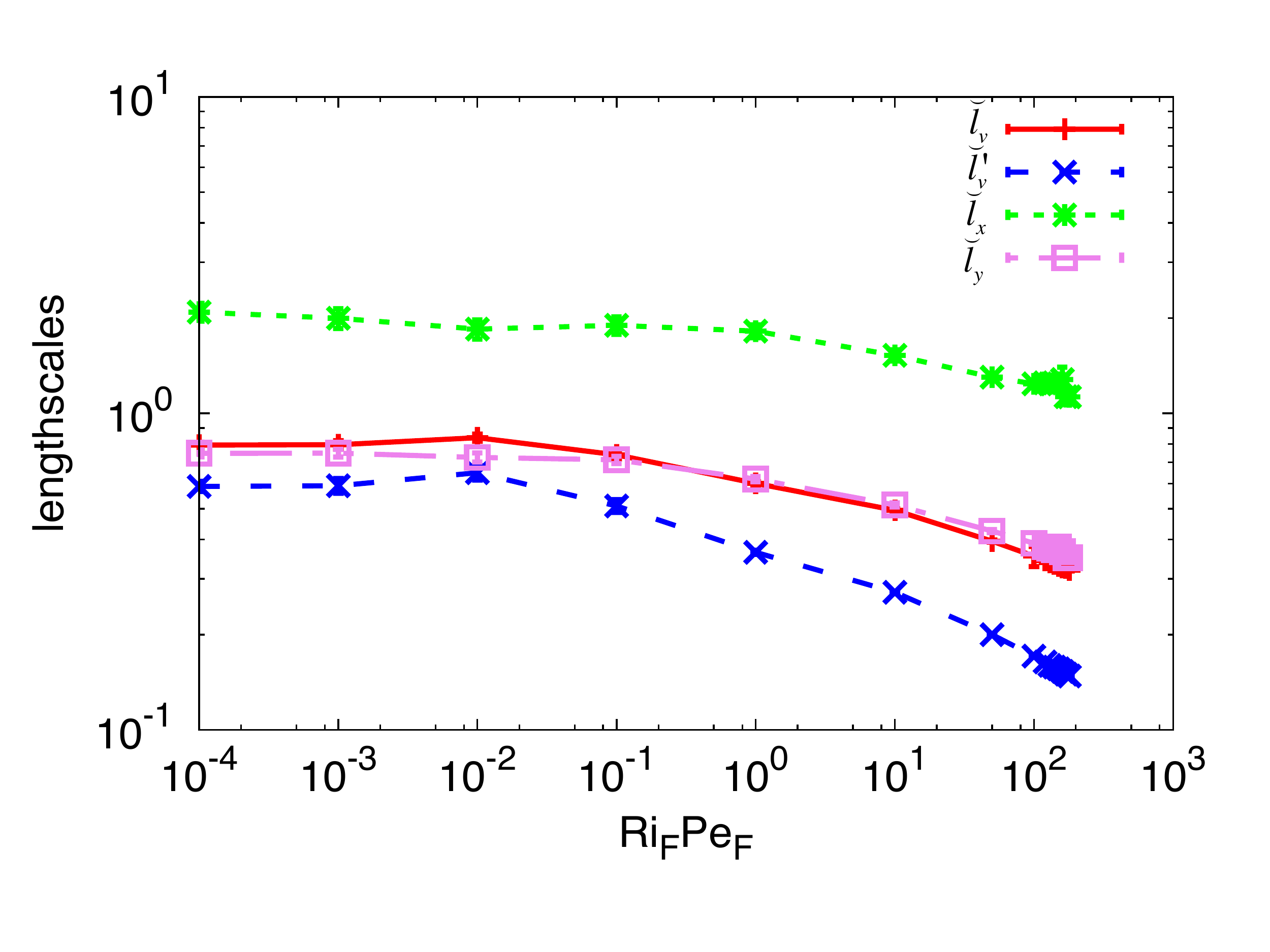}}
  \caption{The vertical length scales $\breve l_v$ and $\breve l_v'$, as well as the horizontal length scales $\breve l_x$ and $\breve l_y$ of the eddies as a function of $\Ri_F \Pe_F$, for $\Re_F = 100$, in the LPN equations (see text for detail).}
\label{fig:lengthscales}
\end{figure}

Figure \ref{fig:lengthscales} shows the variation of $\breve l_x$, $\breve l_y$ and $\breve l_v$ with $\Ri_F\Pe_F$ for $\Re_F = 100$. 
As discussed in Section \ref{sec:qualreslowPe}, $\breve l_v$, as well as $\breve l_x$ and $\breve l_y$, become independent of $\Ri_F\Pe_F$ for sufficiently low $\Ri_F\Pe_F$. The fact that $\breve l_v \rightarrow 0.8 \breve k^{-1}$ at low $\Ri_F \Pe_F$ suggests that the vertical eddy scale is indeed very similar to the imposed forcing lengthscale in this limit, which is not surprising since this is the only available non-diffusive scale in the system when buoyancy is negligible. In the opposite limit, 
we clearly see that the vertical eddy scale gradually decreases as stratification becomes more important. This trend agrees with the common notion that vertical overturning becomes more and more difficult, so only eddies whose vertical scale is small enough for diffusion to mitigate the effect of stratification are allowed. %We find that for $\Ri_F \Pe_F \ge 10$, $l_v \propto (\Ri_F\Pe_F)^{-1/6}$. Note that this scaling is tentative, since it is only based on a few data points. Indeed, the need for $\Ri_F \Pe_F \ge 10$, combined with the fact that we have only found unstable solutions for $\Ri_F \Pe_F \le 180$, means that the range of parameters for which this scaling holds is very narrow (at least for $\Re_F = 100$). However, we shall present in Section \ref{sec:modelexplained}  a simple model that provides a reasonable explanation for the power of -1/6, which gives our experimental findings some theoretical basis. XXX Not sure about this anymore, because this scaling doesn't hold if I use a different definition for lv XXX.
It is interesting, and somewhat surprising, to see that both $\breve l_x$ and $\breve l_y$ also decrease with increasing $\Ri_F \Pe_F$, so the anisotropy of the eddies does not vary as strongly with $\Ri_F\Pe_F$ as one may naively expect: across most of the range, $\breve l_y \simeq \breve l_v$, while $\breve l_x \sim 3.0-3.4 \breve l_v$. In other words, the eddies take the form of ``cigares" of more-or-less circular cross sections in the $(y,z)$ plane, and three times longer in the $x-$direction.  Whether this result still holds at much larger values of $\Ri_F\Pe_F$ remains to be determined.

Since the vertical scale of the eddies is sometimes used \citep[as in][for instance]{Zahn92} to make predictions on the effective mixing rate induced by the turbulence, one would ideally like to create a quantitative model for the variation of $\breve{l}_v$ with $\Ri_F\Pe_F$ and $\Re_F$. However, we first note that the estimated vertical eddy scale depends somewhat on the definition of $\breve{l}_v$ adopted. Indeed, we could equivalently have chosen to use the definition 
\begin{equation}
\breve{l}'_v =  \left[ \frac{\sum_{\breve k_x} \sum_{\breve k_y} \sum_{\breve k_z}  \breve E_w(\breve \bk) \breve k_z   }{\sum_{\breve k_x} \sum_{\breve k_y} \sum_{\breve k_z}  \breve E_w(\breve \bk)  }  \right]^{-1}  \mbox{  ,}
\end{equation}
which is just as plausible as the one advocated in equation (\ref{eq:lv1}). The two definitions would in fact be equivalent (within a constant factor) if the kinetic energy spectrum were a strict power law. However, this is not the case in our simulations, and as a result $\breve{l}_v$ and $\breve{l}_v'$ are not proportional to one another. We find that, for large $\Ri_F\Pe_F$, 
\begin{eqnarray}
\breve{l}_v \sim (\Ri_F \Pe_F)^{-1/6} \mbox{   , } \nonumber \\
\breve{l}'_v \sim (\Ri_F \Pe_F)^{-2/9}  \mbox{   . }
\end{eqnarray}
The fact that the two power laws are different suggests that there are significant changes in the shape of the power spectrum as $\Ri_F \Pe_F$ increases, which in turn implies that it will be difficult to create a simple a priori model for the variation of the eddy scale with stratification. At best, one can look at the empirical data and infer that $\breve{l}_v$ varies {\it approximately} as 
\begin{equation}
\breve{l}_v \sim (\Ri_F \Pe_F)^{-\alpha_l} \mbox{   for  } \Ri_F \Pe_F \gg 1 \mbox{   , }
\end{equation}
where the constant of proportionality is of order one, and where $\alpha_l \simeq 0.19 \pm 0.03$ (depending on the definition adopted). Again, this is a tentative measurement, which would benefit from being confirmed with simulations at higher $\Re_F$ and higher $\Ri_F \Pe_F$, so we caution the reader against using it too far outside of the range of parameters for which the formula has been established.

\subsection{Energy budget and heat transport}
\label{sec:energy}

In order to study the kinetic energy budget, we start from the original set of equations (\ref{eq:originalforcing}), and express them in the non-dimensionalization based on the forcing. We then dot the momentum equation with $\breve \bu$ and multiplying the thermal equation with $\breve{T}$, and take a spatial average over the domain to find that 
\begin{eqnarray}
\frac{1}{2} \frac{\p }{\p t} \langle |\breve \bu|^2 \rangle = \Ri_F \langle \breve w \breve T \rangle - \frac{1}{\Re_F} \langle | \nabla \breve \bu|^2 \rangle + \langle \breve \bu \cdot \breve{\bF} \rangle \mbox{  ,}  \\
\frac{1}{2} \frac{\p}{\p t}\langle \breve T^2 \rangle  = - \langle \breve w \breve T \rangle - \frac{1}{\Pe_F} \langle | \nabla \breve T|^2 \rangle \mbox{  .}
\end{eqnarray}
If, in addition, the system achieves a statistically-stationary state, the thermal energy balance implies:
\begin{equation} 
\langle \breve w \breve T \rangle = - \frac{1}{\Pe_F} \langle | \nabla \breve T|^2 \rangle \mbox{  ,}
\label{eq:thermalmeaneq}
\end{equation}
which shows that the turbulent heat flux $\langle \breve w \breve T \rangle$ caused by shear instabilities is negative (i.e downward) in a stably stratified region. 
Substituting $\langle  \breve w \breve T \rangle$ into the kinetic energy equation, we then have: 
\begin{equation} 
 \langle \breve \bu \cdot \breve \bF \rangle  = \frac{ \Ri_F}{\Pe_F} \langle | \nabla \breve T|^2 \rangle +  \frac{1}{\Re_F} \langle | \nabla  \breve \bu|^2 \rangle \mbox{  ,}
\label{eq:powerdef}
\end{equation} 
which shows how the power input into system by the force $\bF$ is first converted into velocity and temperature fluctuations, which are then both dissipated microscopically. Note that we obtain exactly the same energy balance using the LPN equations, the only difference being that these equations assume that (\ref{eq:thermalmeaneq}) is true at all times instead of in a quasi-stationary, domain-averaged sense (see equation (\ref{eq:slave})). In that case, equation (\ref{eq:powerdef}) can also be rewritten as 
\begin{equation} 
 \langle \breve \bu \cdot \breve \bF \rangle  = \Ri_F\Pe_F \langle | \nabla^{-1} \breve w|^2 \rangle +  \frac{1}{\Re_F} \langle | \nabla  \breve \bu|^2 \rangle \mbox{  .}
\label{eq:powerdef2}
\end{equation} 

\begin{figure}[h]
  \centerline{\includegraphics[width=0.7\textwidth]{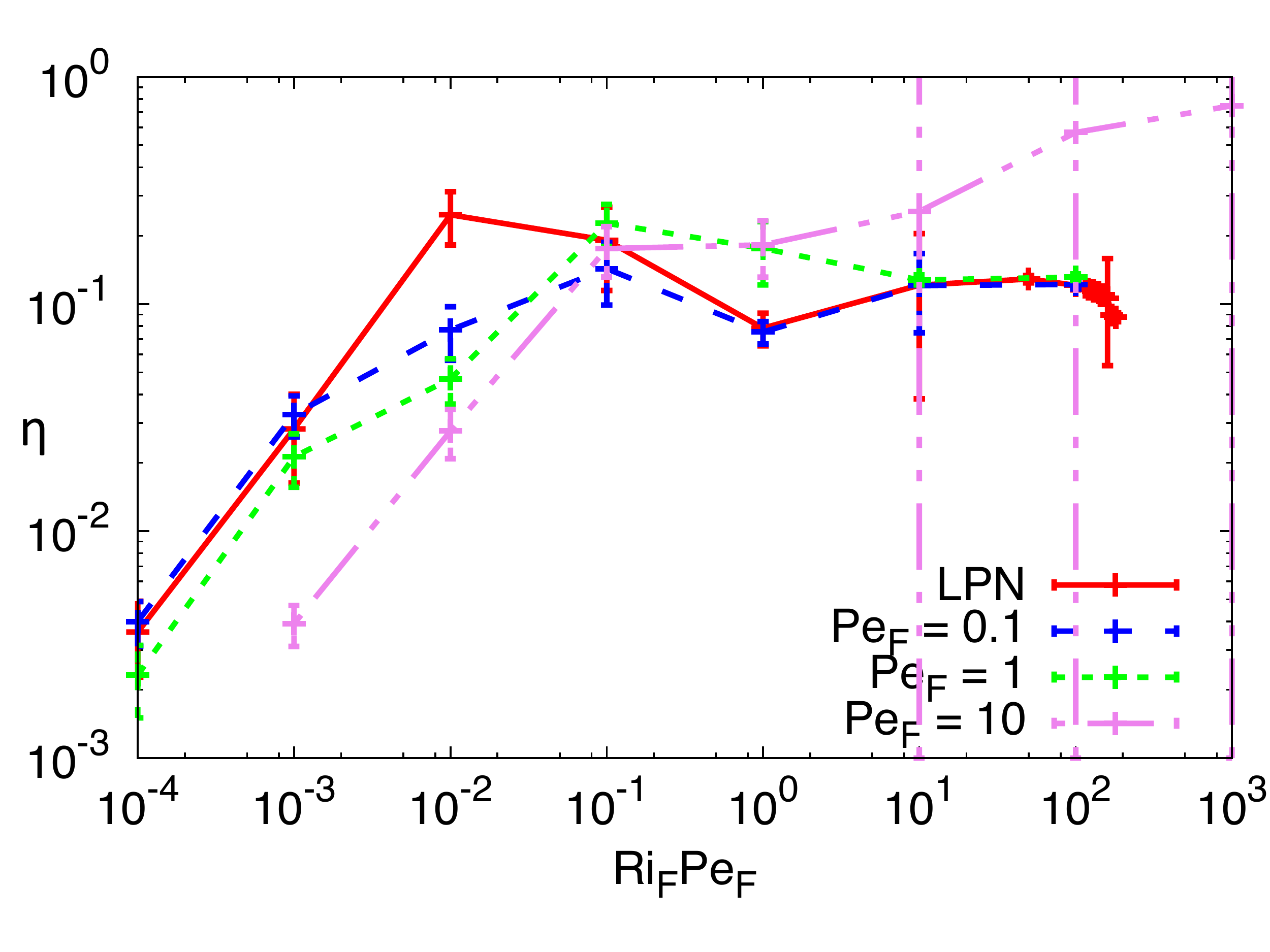}}
  \caption{Buoyancy transport efficiency factor $\eta$ at $\Re_F  = 100$, in the LPN equations and in the standard equations at various $\Pe_F$. }
\label{fig:eta}
\end{figure}

It is common in the geophysical literature to measure the efficiency of stratified turbulence in mixing buoyancy through the ratio  
\begin{equation}
\eta = - \frac{\Ri_F \langle \breve w \breve T \rangle }{\langle \breve \bu \cdot \breve \bF \rangle } = \frac{ \Ri_F \Pe_F \langle | \nabla^{-1} \breve w|^2 \rangle }{\langle \breve \bu \cdot \breve \bF \rangle }  \mbox{ in the LPN limit} 
\end{equation}
which is also called the ``flux Richardson number" \citep{Linden1979}. This quantity measures the fraction of the input power that is effectively used to transport buoyancy while $1-\eta$ measures the fraction of the input power that is dissipated viscously (either through the dissipation of the turbulent fluctuations, or through the dissipation of the mean flow). The ratio $\eta$ is shown in Figure \ref{fig:eta} as a function of $\Ri_F \Pe_F$ for $\Re_F = 100$, and for different values of $\Pe_F$ as well as for the LPN equations. For very low $\Ri_F\Pe_F$, $\eta$ increases linearly with $\Ri_F \Pe_F$ as $\eta \simeq 40 \Ri_F \Pe_F$. This is consistent with the notion that temperature is a passive scalar in that limit, and can be explained mathematically by noting that both $\langle \breve \bu \cdot \breve \bF \rangle$ and $\langle |\nabla^{-1} \breve w|^2 \rangle$ (as with all other quantities in the system) are independent of $\Ri_F \Pe_F$ in effectively unstratified flows. For larger stratifications, we see that in the LPN equations $\eta$ has a  first local maximum between $\Ri_F \Pe_F = 0.01$ and $0.1$, then a dip around $\Ri_F \Pe_F = 1$, then another local maximum later on followed by another dip. While the dip at $\Ri_F \Pe_F = 1$ remains to be explained, the second dip corresponds to the transition to a skewed state, and can be understood by noting that the partitioning of the flow between a turbulent zone and a laminar one  effectively increases the viscous dissipation (in particular, that associated with the mean flow) and reduces the turbulent heat transport.

Generally speaking, however, we see that $\eta$ oscillates around about $0.1$ (for $\Ri_F \Pe_F$ greater than about 0.01), a value which is notably smaller than the typical transport efficiency discussed in the high-Prandtl-number geophysical literature where $\eta$ is typically closer to $0.2$ \citep[see the review by][for instance]{PeltierCaulfield03}. The fact that $\eta$ is smaller at low P\'eclet number presumably stems from the fact that thermal diffusion plays a significant role in dissipating buoyancy fluctuations {\it before} they can cause mixing. Viewed in this light, it is in fact somewhat surprising that $\eta$ is not actually much smaller than $0.1$ in the LPN equations. We also see in Figure  \ref{fig:eta} that the dependence of $\eta$ on $\Ri_F \Pe_F$ is roughly the same for the standard equations at low $\Pe_F$ and for the LPN equations, as expected. For higher $\Pe_F$, however, the results are very different and $\eta$ can be much larger than 0.1 (it can in fact be much larger that 0.2 in the limit of very strongly stratified shear flows).

These results have interesting consequences: they imply that the dimensional heat flux carried by shear-induced turbulence in the low P\'eclet number limit, and in the case where $\Ri_F \Pe_F \ge 0.01$, can be predicted to be 
\begin{eqnarray}
{\cal F} = \rho_0 c_p \left( - \kappa_T \frac{d \bar T}{dr} + \langle wT \rangle \right)  = \rho_0 c_p \left[ - \kappa_T \frac{d \bar T}{dr} + U_F k^{-1} \left(\frac{d\bar T}{dr} - \frac{d T_{\rm ad}}{dr}\right)\langle \breve w \breve T \rangle   \right]    \nonumber \\
= \rho_0 c_p \left[ - \kappa_T \frac{d \bar T}{dr} - \frac{\eta  \langle \breve \bu \cdot \breve \bF \rangle  }{\Ri_F}U_F k^{-1} \left(\frac{d\bar T}{dr} - \frac{d T_{\rm ad}}{dr}\right) \right] = - \rho_0 c_p \left[  \kappa_T \frac{d \bar T}{dr}  + \frac{\eta {\cal  P}}{\alpha \rho_0 g} \right] 
\label{eq:heatflux}
\end{eqnarray} 
where $\rho_0$ is the local background density, $c_p$ is the specific heat at constant pressure, ${\cal P} = \langle \bu \cdot \bF \rangle$ is the dimensional power input into the system, which is always positive at steady-state, $\alpha$ is the thermal expansion coefficient defined in Section \ref{sec:model}, $g$ is the local gravity and $\eta$ is roughly equal to 0.1. The remainder of the power injected into the star, that is, about 90\% of it, is dissipated viscously  and ultimately heats the fluid locally (an effect which is however neglected in the Boussinesq equations).

\subsection{Momentum transport}
\label{sec:momtrans}

The simulations discussed in Sections \ref{sec:num} and \ref{sec:num2} have revealed a number of non-trivial results, in particular when it comes to the shape of the mean flow. Why is the sinusoidal shape preferred for low degrees of stratification, while more strongly stratified shear flows tend to have nearly piecewise linear profiles?  And more curiously, why is there a sudden loss of symmetry in the flow for even more strongly stratified systems? 
%Unfortunately, the exact value of the input power $\langle \bu \cdot \bF \rangle$ (or its non-dimensional version  $\langle \breve \bu \cdot \breve \bF \rangle$)  is not usually known a priori since it depends on the shape and amplitude of the (unknown) mean flow as: 
%\begin{equation}
%\langle  \breve \bu \cdot \breve \bF  \rangle  = \frac{1}{L_z} \int_0^{L_z}  \breve{\overline{u}}(z,t) \sin(z) dz \mbox{  .}
%\label{eq:meanueq}
%\end{equation}
%While the amplitude of the mean shear can in principle be predicted using (\ref{eq:regimes}), on its own it can only yields a very rough order-of-magnitude estimate of $\langle  \breve \bu \cdot \breve \bF  \rangle$, which prompts the question of whether we can do better. 

In order to make progress towards answering these questions, recall that the evolution of the mean flow as a function of time is given by the horizontal average of the horizontal component of the momentum equation, namely
\begin{equation}
\frac{\partial \breve{\bar{u}}}{\partial t} = - \frac{d}{dz} \overline{\breve u \breve w} + \frac{1}{\Re_F} \frac{d^2  \breve{\overline{u}}}{dz^2} + \sin(z) \mbox{  .}
\label{eq:meanueq}
\end{equation}
When the system reaches a statistically stationary state, this equation implies a balance between the divergence of the momentum flux, the forcing, and the mean shear dissipation by viscosity. If the Reynolds stress $\overline{\breve{u}\breve{w}}$ were a known function of the local shear $d\breve{\bar{u}}/dz$ (and of the input parameters $\Ri_F \Pe_F$ and $\Re_F$) then (\ref{eq:meanueq}) could be used in closed form to infer $\breve{\bar{u}}(z)$. Ideally, one would like to use simulations such as the ones presented here to measure the dependence of  $\overline{\breve{u}\breve{w}}$ on the mean shear and the input parameters, and then see if this could illuminate the flow shape problem. However, this type of constant forcing simulation does no lend itself well to this exercise. Indeed, integrating  (\ref{eq:meanueq}) in the statistically stationary state yields (within an additive constant) 
\begin{equation}
  \overline{\breve u \breve w} =  \frac{1}{\Re_F} \frac{d \breve{\bar{u}}}{dz} -  \cos(z) \mbox{   , }
  \end{equation}
  which implies that the Reynolds stress
\begin{equation}
  \overline{\breve u \breve w} \simeq -  \cos(z) + O(\Re_F^{-1}) 
  \end{equation}
becomes nearly independent both $\Ri_F \Pe_F$ and $\Re_F$ for large enough $\Re_F$. This contrived balance implies that momentum transport cannot easily be studied using this approach. We therefore defer the problem of understanding the momentum balance and mean flow shape to a subsequent paper, in which we shall revisit our numerical results in the light of a turbulence closure model (Kulenthirarajah \& Garaud, 2016, in prep.). %In the meantime, we are forced to be satisfied with a mere order-of-magnitude estimate for $\langle  \breve \bu \cdot \breve \bF  \rangle$, as discussed above. 

\subsection{Compositional mixing}
\label{sec:mixing}

We now finally turn to the question of mixing of a passive tracer, and whether the latter is adequately captured by the models of \citet{EndalSofia78} or \citet{Zahn92}.  For this purpose, we have measured, in a limited number of simulations using the standard equations, the vertical flux of a passive tracer by simultaneously solving for the equation
\begin{equation} 
\frac{\partial \breve C}{\partial t} + \breve \bu \cdot \nabla \breve C + \breve w = \frac{1}{\Pe_C} \nabla^2 \breve C \mbox{   , }
\end{equation}
where $C$ represents for instance the concentration of a particular chemical species. This equation is already non-dimensionalized using $[C] =  k^{-1} \frac{d\bar C}{dr} $ where $ \frac{d\bar C}{dr} $ is an assumed constant dimensional background compositional gradient. The momentum equation remains unchanged. The scalar P\'eclet number $\Pe_C$ is defined as
\begin{equation}
\Pe_C = \frac{\kappa_T}{\kappa_C}\Pe_F  \mbox{   . }
\end{equation}
In all cases we have used $\Pe_C = 100$, so the dependence of the results on $\Pe_C$ have not been tested yet. Note that none of the mixing models currently discussed have any dependence on $\Pe_C$. 

Using these simulations, we extract the rate of scalar mixing by measuring the compositional flux $\langle \breve w \breve C \rangle$. An effective turbulent diffusion coefficient can then be computed via
\begin{equation}
 \langle w C \rangle = -  D\frac{d\bar C}{dr}  \leftrightarrow \breve{D} \equiv \frac{kD}{U_F} =  - \langle \breve w \breve C \rangle \mbox{   . }
 \end{equation}
 
\begin{figure}
 \centerline{\includegraphics[width=\textwidth]{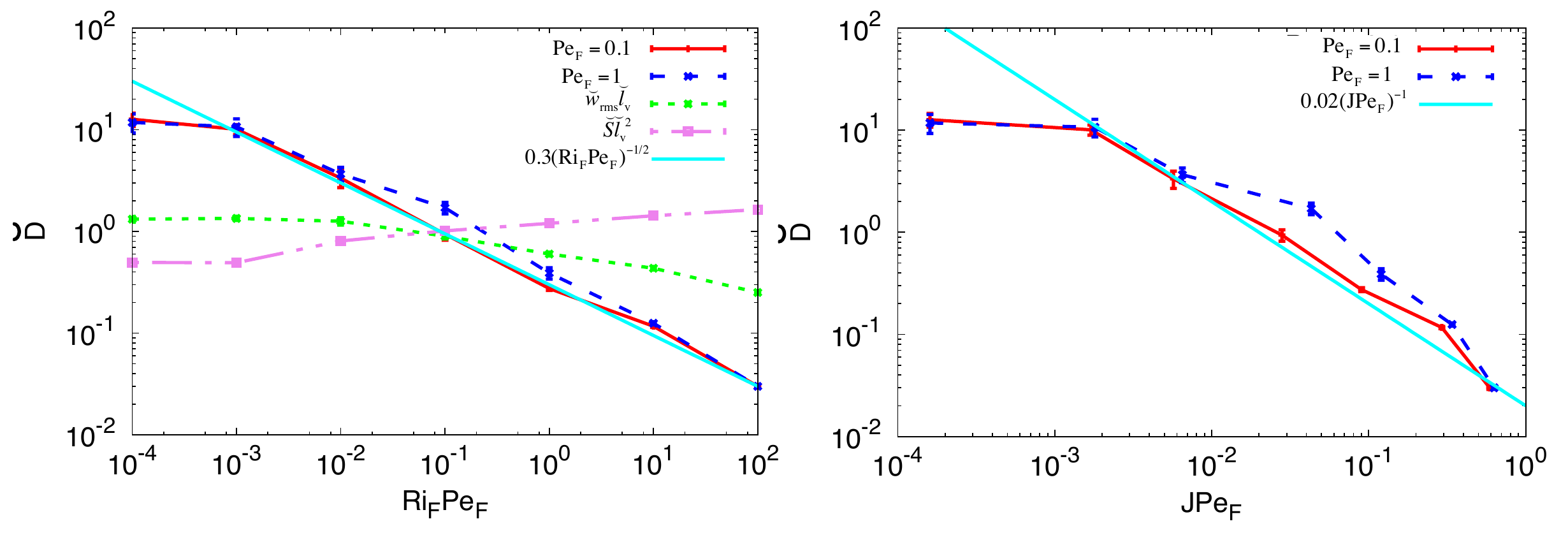}}
 \caption{Left: Turbulent diffusion coefficient $\breve D$ as a function of  $\Ri_F \Pe_F$, for $\Pe_F = 0.1$ and $\Pe_F = 1$. In low P\'eclet number flows, $\breve D$ appears to be well-approximated by $0.3(\Ri_F \Pe_F)^{-1/2}$ for $\Ri_F\Pe_F> 0.001$. Also shown are standard estimates of $\breve D$ as $\breve w_{\rm rms} \breve l_v$ and $\breve S \breve l_v^2$, neither of which are good. Right: Validation of Zahn's model, which predicts that $\breve D \sim (J \Pe_F)^{-1}$. We find that this is the case for $J \ge 0.002$.}
\label{fig:scalartransport}
\end{figure}
 
Our results are shown in Figures \ref{fig:scalartransport}a and \ref{fig:scalartransport}b. Figure \ref{fig:scalartransport}a shows $\breve D$ as a function of the input parameter $\Ri_F \Pe_F$ for simulations with $\Re_F = 100$ and two different low P\'eclet numbers (note that we do not have compositional data for the LPN runs). As expected, the two datasets are reasonably consistent with one another.  In the nearly unstratified limit ($\Ri_F \Pe_F \le 0.001$), we see that $\breve D$ tends to a constant that is roughly equal to 10. For larger values of $\Ri_F \Pe_F$, $\breve D$ appears to follow a power law with $\breve D \simeq 0.3 (\Ri_F \Pe_F)^{-1/2}$. This scaling, as we shall see below, is consistent with Zahn's model, and appears to be valid from $\Ri_F \Pe_F \simeq 0.01$ to  $\Ri_F \Pe_F \simeq 100$. Unfortunately, we do not have any data for larger $\Ri_F \Pe_F$, which would correspond to the very strongly stratified limit. 

It is interesting to note, however, that the variation of $\breve D$ with $\Ri_F \Pe_F$ cannot be explained either by assuming that $\breve D \propto \breve w_{\rm rms} \breve l_v$ (i..e the product of a typical vertical velocity with a typical turbulent lengthscale), nor by assuming that $\breve D \propto \breve S \breve l_v^2$ (where $\breve S$ is the non-dimensional mean shear at mid-layer, see Equation \ref{eq:gradRich}). By contrast with $\breve D$, the former decreases much more weakly with $\Ri_F \Pe_F$, while the latter actually increases slightly with $\Ri_F \Pe_F$. This shows that using these standard estimates for $\breve D$ can be very misleading.

Figure \ref{fig:scalartransport}b shows the same data for $\breve D$, but this time plotted against $J \Pe_F$. We find in this case that $\breve D \simeq 10$ for $J \Pe_F \le 0.002$, while $\breve D \simeq 0.02 (J \Pe_F)^{-1}$ for $J \Pe_F \ge 0.002$. The scaling for $\breve D$ in that limit is therefore consistent with the model of \citet{Zahn92}, and suggests that, dimensionally speaking,
\begin{eqnarray}
&& D \simeq 10 k^{-1} U_F  \mbox{   for  }  J \Pe_F < 0.002\mbox{   , }  \nonumber \\
&& D \simeq 0.02 \frac{ \kappa_T }{J }  \mbox{ otherwise} \mbox{   , }
\end{eqnarray}
where $J = N^2 / S^2 = \Ri_F / \breve S^2$. The reason why Zahn's model appears to explain our data is, however, obscure. As discussed in Section \ref{sec:intro}, \citet{Zahn92} assumes that $D \propto Sl^2$, where $l$ is the largest possible eddy size for which the shear is still nonlinearly unstable according to the criterion $J \Pe_l \le (J \Pe)_c$. As seen in Figure \ref{fig:scalartransport}a, however, that lengthscale cannot be $l_v$ (since $S l_v^2$ is a very poor model for $D$). This leaves us in the rather uncomfortable position of either trying to explain why the lengthscale that dominates the transport of passive scalars should be so very different from the vertical lengthscale of the energy-bearing eddies -- something that goes against what is commonly assumed, or, to accept that the good match between our data and Zahn's model is somewhat of a coincidence. Neither of these options are particularly satisfactory but both are equally plausible (or implausible). Future simulations using a different model setup will be needed to resolve this frustrating conundrum.

\section{Summary and discussion}
\label{sec:ccl}

\subsection{Summary of our findings}

In this paper, we have studied the dynamics of shear instabilities in stably stratified stellar regions, in the limit where the ratio of the thermal diffusion timescale to the turnover timescale of turbulent eddies is short (the so-called ``low P\'eclet number" limit). We have shown that this limit could be relevant in the envelopes of very massive stars, where the thermal diffusivity is in excess of $10^{14}$cm$^2$/s, but most likely does not apply for lower-mass stars, or deep within the interiors of massive stars (see Section \ref{sec:stars}). Low P\'eclet number shear layers can be formally studied with the LPN equations derived by \citet{Lignieres1999}, which can be integrated numerically faster than the standard equations (at least in spectral codes), see Section \ref{sec:model}. 

In this first study, we have chosen to ignore the possibility of compositional stratification and horizontal shear, and focussed only on the case of thermal stratification with vertical shear. We also specifically investigated the dynamics of shear flows that arise from the application of a constant-amplitude, spatially-periodic body force. In this setup, the mean shear is not prescribed, but instead is one of the quantities that we measure. From this modeling choice, three numbers naturally emerge: $\Re_F$, $\Pe_F$ and $\Ri_F$ defined in equation (\ref{eq:forcingnumbers}), which we have shown to be good approximations to the actual turbulent Reynolds number, turbulent P\'eclet number and turbulent Richardson numbers of the resulting statistically-stationary shear flow respectively. As such, they ideally characterize the dynamics of the shear. We first found that while the LPN equations are only formally valid  in the limit $\Pe_F \rightarrow 0$, they are already a good-to-excellent approximation of the full equations whenever $\Pe_F \le 1$. In this limit, the only relevant non-dimensional parameters are the Richardson-P\'eclet product $\Ri_F \Pe_F$ and the Reynolds number $\Re_F$. As such, instability is possible for large $\Ri_F$ as long as $\Pe_F$ is small enough. 

As shown by \citet{Garaudal15}, for large enough Reynolds number this sinusoidal shear flow is linearly unstable provided $\Ri_F \Pe_F \le 0.25\Re_F$, and is energy-stable (i.e. stable to any possible perturbations of any amplitude) for $\Ri_F \Pe_F$ greater than $\xi \Re_F^3$  where $\xi$ is of order one. In practice, finite amplitude instabilities are rather difficult to trigger, and have only been found up to $\Ri_F \Pe_F$ of the order of a few times $\Re_F$, which is not much larger than the linear instability threshold itself. Whether the turbulent solution can be continued for larger $\Ri_F \Pe_F$ remains to be determined. If these results can be directly applied to stars, they strongly suggest that solutions in the high $\Ri_F\Pe_F$ linearly stable/nonlinearly unstable region of parameter space can only be found if the shear slowly and progressively decreases over time from a point when it was linearly unstable, or if the stratification progressively increases likewise. In the opposite case, the shear flow would start as a laminar flow and remain so until the conditions are such that it becomes linearly unstable. 
In practice, this implies that whether a shear flow in the linearly stable/nonlinear unstable region of parameter space is actually turbulent or not depends more on its history than on the present conditions, a classical case of hysteresis. 

Using DNS, we have found that the dynamics of low P\'eclet number shear flows can be divided into three categories. In the limit where $\Ri_F \Pe_F \le 0.01$, the stratification has a negligible effect on the flow dynamics, and temperature behaves as a passive tracer. The amplitude of the resulting mean shear is of order $S \sim (kF_0/\rho_0)^{1/2}$ where $F_0$ is the amplitude of the forcing, $k^{-1}$ is the typical lengthscale of the forcing, and $\rho_0$ is the local density. Most of the kinetic energy, however, is in the turbulent fluctuations rather than in the mean flow. Heat is transported downward, as expected from turbulence in stably stratified fluids. The turbulent heat flux can be derived from (\ref{eq:heatflux}) with $\eta \simeq 40 \Ri_F \Pe_F$, and is dimensionally proportional to the local stratification as measured by $N^2$. The effective diffusivity of a passive tracer is given by $D \simeq 10  (F_0/\rho_0 k^3)^{1/2}$. 

In the opposite limit, where $1 \ll \Ri_F \Pe_F < (\Ri_F \Pe_F)_c$ (the exact lower and upper thresholds depend on the Reynolds number, as described in equation \ref{eq:regimes}), the stratification entirely governs the dynamics of the shear flow. This region of parameter space is usually linearly stable. Most of the total kinetic energy now lies in the mean flow, and the mean shear adjusts itself to be in a state of marginal nonlinear stability, which is captured by a criterion similar to Zahn's criterion \citep[][see Equation \ref{eq:jzahn}]{Zahn1974}, namely $J \Pr  \simeq 0.006$ where $\Pr$ is the Prandtl number and $J  = N^2  / S^2 $ is the local gradient Richardson number. This implies that the mean shear is simply proportional to $N$ as $S \simeq  ( \Pr / 0.006)^{1/2}  N$ in this limit. Meanwhile the turbulent heat flux is given by  (\ref{eq:heatflux}) with $\eta \simeq 0.1$, which physically means that a universal constant fraction of the total input power provided by the forcing, namely 10\%, goes into buoyancy transport while the rest is dissipated viscously.  

Finally, in the intermediate regime, the variation of the mean shear with stratification is more difficult to interpret (see Section \ref{sec:mean}), with $J \Pe_F \simeq 0.1 (\Ri_F \Pe_F)^{1/2}$. The turbulent heat flux is however still reasonably well approximated by  (\ref{eq:heatflux}) with $\eta \simeq 0.1$. In both the intermediate and in the strongly stratified regime, we find that the effective diffusivity of a passive tracer satisfies Zahn's mixing model \citep{Zahn92} with $D \simeq 0.02 \kappa_T /J$, even though the reason why this is the case remains elusive (see Section \ref{sec:mixing} for detail). In particular, we have found that contrary to what is commonly assumed in astrophysics, $D$ is not well-approximated by the product of the typical vertical velocity and typical lengthscale of the turbulent eddies. 

Finally, we have found that for large enough $\Re_F$ and for $\Ri_F \Pe_F$ approaching the nonlinear stability threshold $(\Ri_F \Pe_F)_c$, the system spontaneously transitions into new state which does not have the same symmetries as the imposed forcing. Instead, the mean shear becomes skewed and is partitioned between wider turbulent regions of moderate shear, and thinner laminar regions of very strong shear. The reason for this transition remains to be determined, but could be attributed to the non-monotonicity of the stress-strain relationship, in a manner that is similar to the spontaneous formation of layers in stratified systems that have non-monotonous relationships between the buoyancy flux and the buoyancy gradient \citep{ballsy}. In order to study this effect in more detail, and confirm that it is a generic result rather than a peculiarity of our sinusoidal forcing assumption, we shall need to use a different model setup which is not as strongly constrained in terms of the momentum transport balance as the one we are currently using (see Section \ref{sec:momtrans} and below for more detail). This will be the subject of a forthcoming study. 

\subsection{Discussion and outlook}

All of the aforementioned results have been obtained for simulations with moderate turbulent Reynolds numbers $\Re_F$ up to 100, but need to be confirmed with future runs at higher $\Re_F$ when these become feasible in a more reasonable amount of time than what is currently possible. Specifically, we need to better determine (a) the upper limit for instability to finite amplitude perturbations $(\Ri_F \Pe_F)_c$ as a function of $\Re_F$ and (b) the variation with $\Re_F$ of the various dynamical regime thresholds discussed in Section \ref{sec:mean} and the dependence of the mean shear, heat flux, and turbulent diffusivity on $\Ri_F \Pe_F$ within each regime. 

We have, on the other hand, run a number of simulations with different domain sizes to determine the impact of the system geometry. Generally-speaking, all of the simulations presented here were done in a domain that is sufficiently large to guarantee that halving it does not affect the results much, and in the few cases where we have doubled the domain width, length or height, no difference in the mean quantities larger than the error-bars quoted was noticed. The only exception was in the case of the skewed runs when the domain height was doubled. In that case, the system exhibits interesting time-dependent  dynamics, whereby the mean shear oscillates between different skewed quasi-steady states. However, given our reservations about the contrived nature of this type of forcing, we defer a discussion of these results until the momentum transport through the system is better understood (see Kulenthirarajah \& Garaud, 2016, in prep.). 

Indeed, while our choice of using a sinusoidal forcing was principally motivated by numerical convenience, this kind of model appears to suffer from two distinct issues. First, as discussed in Section \ref{sec:momtrans}, the imposed forcing strongly constrains the horizontal momentum transport equation, to the extent that it is difficult to study momentum transport with this setup. As a result, what controls the overall shape and amplitude of the mean flow still remains, to a great extent, unknown. More crucially, as noted by \citet{Garaudal15}, the dependence of the threshold for linear stability on $\Re_F$ is fundamentally different in the case of a sinusoidal shear and in the case of a hyperbolic tangent shear layer \citep{Lignieresetal1999}, the latter being much closer to the energy stability threshold than the former. This suggests that the shape of the shear itself plays a crucial role in the development of shear instabilities. Hence, how much of the results obtained here are specific to the sinusoidal shear case, and how much are generic to all types of shear layers, remains to be determined. 

There are several alternatives to using a sinusoidal forcing. A commonly adopted solution is to use a uniform shear model \citep{Rogallo81,Jacobitzal97,Bruckeral07,MatheouChung12,ChungMatheou12,PratLignieres13,PratLignieres14}. In this case, the background shear is somehow imposed and maintained, and the turbulence merely adjusts itself to the given shearing rate. There are two difficulties associated with this approach, the first being the maintenance of the shear. The simplest way to create a background linear shear flow is to use, as in \citet{PratLignieres13}, an adaptive force which is calculated at every timestep to drive the shear precisely back towards a linear profile should it start deviating away from it. A disadvantage of this model is that the energetics of the system are difficult to study, since there is no control over the force actually required to maintain the shear. A second possibility is to use the shearing sheet model, well-known in the context of astrophysical disks. The numerical implementation of this approach is, however, more difficult especially in a spectral code \citep[see the discussions in][]{Bruckeral07}. 

The second difficulty with using a constant shear background flow is the fact that this system is well-known to be always linearly stable \citep{Knobloch84}, whether unstratified or stratified, and requires finite amplitude perturbations to trigger and maintain turbulence. This, in itself, is not a technically difficult problem to overcome -- one simply needs to find the unstable finite amplitude branch of solution for relatively low $\Ri_F \Pe_F$ and then follow it by continuation for higher and higher $\Ri_F \Pe_F$, as we have done already in this paper. However, it is also the case that even a small amount of curvature in the shear, together with the presence of an inflection point, is all that is needed to have instability to infinitesimal perturbations (for low enough $\Ri_F \Pe_F$). This raises an interesting question: all parameters being the same, is the nature of the turbulence significantly different when triggered by supercritical instabilities than when triggered by subcritical ones? In fact, there is no reason to believe that this would be the case. Even far from the onset of instability, the shape and growth timescale of the fastest-growing linearly unstable modes can still influence the dynamics of a fully turbulent flow, serving as its injection scale. Meanwhile, the same statement could perhaps apply to the finite amplitude subcritical modes, but these are likely to be very different in nature from the global linear modes. In other words, we suspect that the turbulent dynamics obtained in a simulation at constant shear are not always necessarily representative of the ones one may obtain from global simulations when linear instabilities are present. Clearly, both approaches (spatially varying body force or constant background shear) have their pros and cons, and it is our belief that much can be learned from comparing the outcomes of both types of simulations run at similar parameters. This will be the subject of future work.

\acknowledgements

This work is funded by NSF-AST1517927 and NSF-AST1412951. The authors thank Stephan Stellmach for granting us the use of his excellent code, and for many useful discussions. The authors also thank Neil Balmforth, Tobias Bischoff, Nic Brummell, Colm-cille Caulfield,  Basile Gallet and Paul Linden for their insight into this work. The simulations presented in this paper were run on the Hyades cluster at UC Santa Cruz, purchased using an NSF-MRI grant. This paper is written in memory of Jean-Paul Zahn, whose visionary work continues to inspire us today.

%\bibliographystyle{apj}
%\bibliography{NSF-bib}
% Note the spaces between the initials
\providecommand{\noopsort}[1]{}\providecommand{\singleletter}[1]{#1}%

\end{document}